\documentclass[12pt,preprint]{aastex}
\shorttitle{Zodiacal Cloud Origin}
\shortauthors{Nesvorn\'y et al.}
\begin{document}
\baselineskip 19.pt
\title{\normalsize Cometary Origin of the Zodiacal Cloud and Carbonaceous Micrometeorites. Implications 
for Hot Debris Disks}
\author{David Nesvorn\'y$^{1,2}$, Peter Jenniskens$^3$, Harold F. Levison$^{1,2}$, William F. Bottke$^{1,2}$,
\and David Vokrouhlick\'y$^4$} 
\affil{(1) Department of Space Studies, Southwest Research Institute,  
1050 Walnut St., Suite~400, Boulder, Colorado 80302, USA}
\affil{(2) Center for Lunar Origin \& Evolution, NASA Lunar Science Institute, Boulder, Colorado 80302, USA}
\affil{(3) Carl Sagan Center, SETI Institute, 515 N. Whisman Road, Mountain View, CA 94043, 
USA}
\affil{(4) Institute of Astronomy, Charles University, V Hole\v{s}ovi\v{c}k\'ach 2, 180 00 Prague 
8, Czech Republic}
\begin{abstract}

The zodiacal cloud is a thick circumsolar disk of small debris particles produced by asteroid 
collisions and comets. Their relative contribution and how particles of different sizes dynamically evolve 
to produce the observed phenomena of light scattering, thermal emission, and meteoroid impacts are unknown. 
Until now, zodiacal cloud models have been phenomenological in nature, composed of ad-hoc components with 
properties not understood from basic physical processes. Here, we present a zodiacal cloud model based 
on the orbital properties and lifetimes of comets and asteroids, and on the dynamical evolution of 
dust after ejection. The model is quantitatively constrained by IRAS observations of thermal emission, 
but also qualitatively consistent with other zodiacal cloud observations, with meteor observations, 
with spacecraft impact experiments, and with properties of recovered micrometeorites. We find that 85-95\% 
of the observed mid-infrared emission is produced by particles from the Jupiter-family comets (JFCs) 
and $<$10\% by dust from long period comets. The JFC particles that contribute to the observed cross-section 
area of the zodiacal cloud are typically $D\approx100$ $\mu$m in diameter. Asteroidal dust is found to be 
present at $<$10\%. We suggest that spontaneous disruptions of JFCs, rather than the usual cometary activity driven by 
sublimating volatiles, is the main mechanism that librates cometary particles into the zodiacal cloud. 
The ejected mm to cm-sized particles, which may constitute the basic grain size in comets, are disrupted 
on $\lesssim$10,000 yrs to produce the 10-1000 $\mu$m grains that dominate the thermal emission and mass
influx. Breakup products with $D>100$~$\mu$m undergo a further collisional cascade with smaller
fragments being progressively more affected by Poynting-Robertson (PR) drag. Upon reaching $D<100$~$\mu$m,
the particles typically drift down to $<$1 AU without suffering further disruptions. The resulting 
Earth impact speed and direction of JFC particles is a strong function of particle size. While 300 $\mu$m
to 1 mm sporadic meteoroids are still on eccentric JFC-like orbits and impact from antihelion/helion 
directions, which is consistent with the aperture radar observations, the 10-300 $\mu$m particles have 
their orbits circularized by PR drag, impact at low speeds and are not detected by radar. Our results 
imply that JFC particles represent $\sim$85\% of the total mass influx at Earth. Since their atmospheric 
entry speeds are typically low ($\approx$14.5 km s$^{-1}$ mean for $D=100$-200~$\mu$m with $\approx$12 km 
s$^{-1}$ being the most common case), many JFC grains should survive frictional heating and land on the 
Earth's surface. This explains why most micrometeorites collected in antarctic ice have primitive 
carbonaceous composition. The present mass of the inner zodiacal cloud at $<$5 AU is estimated to be 
1-$2\times10^{19}$ g, mainly in $D=100$-200~$\mu$m particles. The inner zodiacal cloud should have 
been $>$10$^4$ times brighter during the Late Heavy Bombardment (LHB) epoch $\approx$3.8 Gyr ago, when 
the outer planets scattered numerous comets into the inner solar system. The bright debris disks with a 
large 24-$\mu$m excess observed around mature stars may be an indication of massive cometary populations 
existing in those systems. We estimate that $\sim$$10^{22}$, $\sim$$2\times10^{21}$ and $\sim$$2\times10^{20}$ 
g of primitive dark dust material could have been accreted during LHB by the Earth, Mars and Moon, 
respectively.
\end{abstract}
\section{Introduction}
The zodiacal cloud is a dynamic assembly of meteoroids in bound orbits around the Sun. The orbits depend on 
particle size, location in the cloud, and the type of parent body. Interstellar dust particles that pass 
through the solar system are not considered in this paper, nor are small meteoroid fragments that move out 
of the solar system on hyperbolic orbits (``beta-meteoroids''). 

Traditionally, the zodiacal cloud has been described with phenomenological models of dust distributions to 
explain the amount of scattered light (Hong, 1985; Kniessel and Mann, 1991; Ishiguro et al. 1999; Hahn 
et al., 2002), the Doppler shifts of the solar Mg I Fraunhofer line (Hirschi and Beard, 1987, Mukai and Mann, 
1993; Clarke et al., 1996; Reynolds et al., 2004), and the more easily to interpret thermal emission observed in 
various lines of sight (Kelsall et al., 1998; Maris et al., 2006). Particularly good 
scattered light observations came from the Clementine mission (Hahn et al., 2002), while thermal infrared 
observations are mostly from the Infrared Astronomical Satellite - IRAS (Low et al., 1984; Hauser et al., 1984; 
Good et al., 1986; Sykes et al., 1990), the Cosmic Background Explorer - COBE (Reach et al., 1995; Kelsall et 
al., 1998), the Midcourse Space Experiment - MSX (Price et al., 2003), the Infrared Space Observatory - ISO 
(Fixsen and Dwek, 2002; Leinert et al., 2002; Reach et al., 2003; Mueller et al. 2005), and the Spitzer Space 
Telescope (Bhattacharya and Reach, 2004; Reach et al., 2007). 

The femenological models successfully describe the size, spatial and velocity distributions of dust particles 
in the solar system (e.g., Gr\"un et al., 1985; Divine, 1993; Dikarev et al., 2004). They are particularly useful 
for accessing the satellite impact hazard, designing spacecraft impact experiments and studies of extrasolar 
emission sources such as the cosmic microwave background (e.g., Kelsall et al., 1998). The femenological models, 
however, fall short in answering basic questions related to the origin of the zodiacal cloud, its temporal 
brightness variability, and the provenance of interplanetary particles collected at the Earth. Consequently, 
the origin of the zodiacal cloud, interplanetary dust particles (IDPs) collected in the Earth's stratosphere 
(e.g., Love and Brownlee, 1994) and micrometeorites (MMs) on the ground (Taylor et al., 1996; Engrand and 
Maurette, 1998; Farley et al., 1998, 2006; Genge, 2006) is still being a matter of considerable debate. This 
limits our ability to link the detailed laboratory studies of IDPs and MMs to the properties of their parent 
bodies, and to use the zodiacal cloud as a valuable reference for studies of the exozodiacal debris disks. 

Detailed dynamical models can be more useful in this context. At the root of dynamical models are the physical 
properties of interplanetary dust, such as density, geometric albedo, elemental composition, mineralogy, tensile 
strengh, heat capacity, etc. (e.g., Dumont and Levasseur-Regourd, 1988; McDonnell and Gardner, 1998; Gustafson, 1994; 
Gustafson et al., 2001; Levasseur-Regourd et al., 2001), which determine the behavior of particles in interplanetary 
space (e.g., planetary perturbations, collisions, sublimation, sputtering) and their interaction with a detector 
(e.g., ablation of micrometeorites in the Earth's atmosphere, $^3$He retention, thermal radiation, light scattering). 
In dynamical models, the individual particles are tracked by numerical codes as they evolve by various processes 
from their sources (assumed to be, e.g., asteroids, comets, satellites or Kuiper belt objects) to sinks (e.g., 
when they sublimate, disrupt, impact or leave the solar system). Insights into the origin of the zodiacal cloud 
can be obtained by calibrating the results of dynamical models on observations. 

Until now, detailed dynamical models have been only developed for asteroidal dust to explain the origin of the zodiacal 
dust bands (e.g., Dermott et al. 1984; Grogan et al., 1997, 2001; Reach et al., 1997; Nesvorn\'y et al. 2006; 
Vokrouhlick\'y et al., 2008) and trapped dust in Earth's Langrange points first seen in IRAS observations (Dermott et al., 1994a).
It has been established that the dust bands originate from the youngest asteroid families (Nesvorn\'y et al., 2003, 
2008). However, claims that asteroids are a major if not dominant source of zodiacal dust, by assuming that all 
main belt asteroids contribute dust (e.g., Dermott et al. 1995; Durda and Dermott, 1997; Kortenkamp and Dermott, 
1998) has remained in doubt. 

Models of the zodiacal cloud need not only explain line-of-sight properties, but also the observed 
influx of meteors (see Ceplecha et al., 1998; Jenniskens, 2006, for a review) and the impact 
rate of meteoroids on satellites (Love and Brownlee, 1993). Until now, models that were developed to explain these 
dynamical phenomena (e.g., Gr\"un et al., 1985; Divine et al., 1993; Kessler et al. 1994; Staubach et al., 1997; 
Dikarev et al., 2004) were based on ad-hoc populations of meteoroids in various types of orbits without a 
dynamical underpinning to sources and sinks. Moreover, 
all current satellite impact models use meteoroid velocity distributions (both magnitude and spatial direction) 
derived from meteor observations (Taylor and Elford, 1998; Jones and Brown, 1993; Brown and Jones, 1999; Brown 
and Campbell-Brown, 2003), which pertain to much bigger particles than typically encountered by satellites. 

In this paper, we investigate what fraction of the zodiacal cloud is due to cometary versus asteroidal dust 
by calculating the evolution of dust particles under solar radiation forces and planetary perturbations (including 
resonances and close encounters), ejected from model populations of all potential sources (not just representative 
examples). The source populations include asteroids, active and mostly dormant Jupiter-family comets (JFCs), 
Halley-type comets and long-period Oort-cloud comets. In recent years, much insight was gained 
into the dynamical characteristics of these populations and the number of asteroids and comets 
that can contribute dust to the zodiacal cloud (e.g., Levison and Duncan, 1994, 1997; Jedicke and Metcalfe, 1998; 
Wiegert and Tremaine, 1999; Dones et al., 2004; Francis, 2005; Gladman et al. 2009). At the same time, it was 
realised that mostly dormant JFCs are the main source of meteoroid streams in the inner solar system (Jenniskens 
2006; 2008) and responsible for the antihelion/helion sources in the sporadic meteoroid background (Jenniskens, 
2006; Wiegert et al., 2009). 

Here we couple these new insights to dynamical behaviour of cometary and asteroidal dust in order to evaluate the 
contribution of various sources to the thermal emission of zodiacal dust and the influx of micrometeorites. We show 
that JFCs are the main source of zodiacal dust inside 5 AU and the most likely source of the micrometeorites found 
on Earth. Our results also provide quantitative constraints on dust lifetimes, influx rates, and velocity distributions 
directly from the known abundances of meteoroid parent bodies. The results are used to quantify the properties of 
zodiacal dust cloud in the past and discuss implications for studies of exozodiacal debris disks. 
 
To set up the stage for our modeling described in \S3, we discuss IRAS observations of the zodiacal 
cloud in \S2. Results are presented in \S4. In \S5, we estimate the current and historical terrestrial accretion
rates of dust and discuss the implications of our work for studies of micrometeorites and debris disks. Comet 
disruptions/splitting events are reviewed in \S6. We suggest that they are the main mechanism by which the cometary 
particles are liberated from their parent bodies into the zodiacal cloud. Previous work and origin of dust particles 
beyond Jupiter are discussed in \S7 and \S8, respectively. 
\section{Constraints}
Our primary constraints are the observations of the zodiacal cloud by IRAS which
have been confirmed by COBE and Spitzer (Hauser et al., 1984; Low et al., 1984; Kelsall et al., 1998;
Sykes et al., 2004). IRAS measured mid-infrared (MIR) fluxes in four filters with effective 
wavelengths of 12, 25, 60 and 100~$\mu$m.  These filters can be used as windows into 
the dust distribution at different distances from the Sun. Measurements in the 12-$\mu$m 
IRAS band are mainly sensitive to distributions of particles at $\sim$1-2 AU, while the 25- 
and 60-$\mu$m band measurements preferentially detect thermal emission from larger 
distances. IRAS observations in the 100-$\mu$m band are less useful for probing the 
thermal radiation of dust particles in the inner solar system.

\begin{figure}
\epsscale{0.8}
\plotone{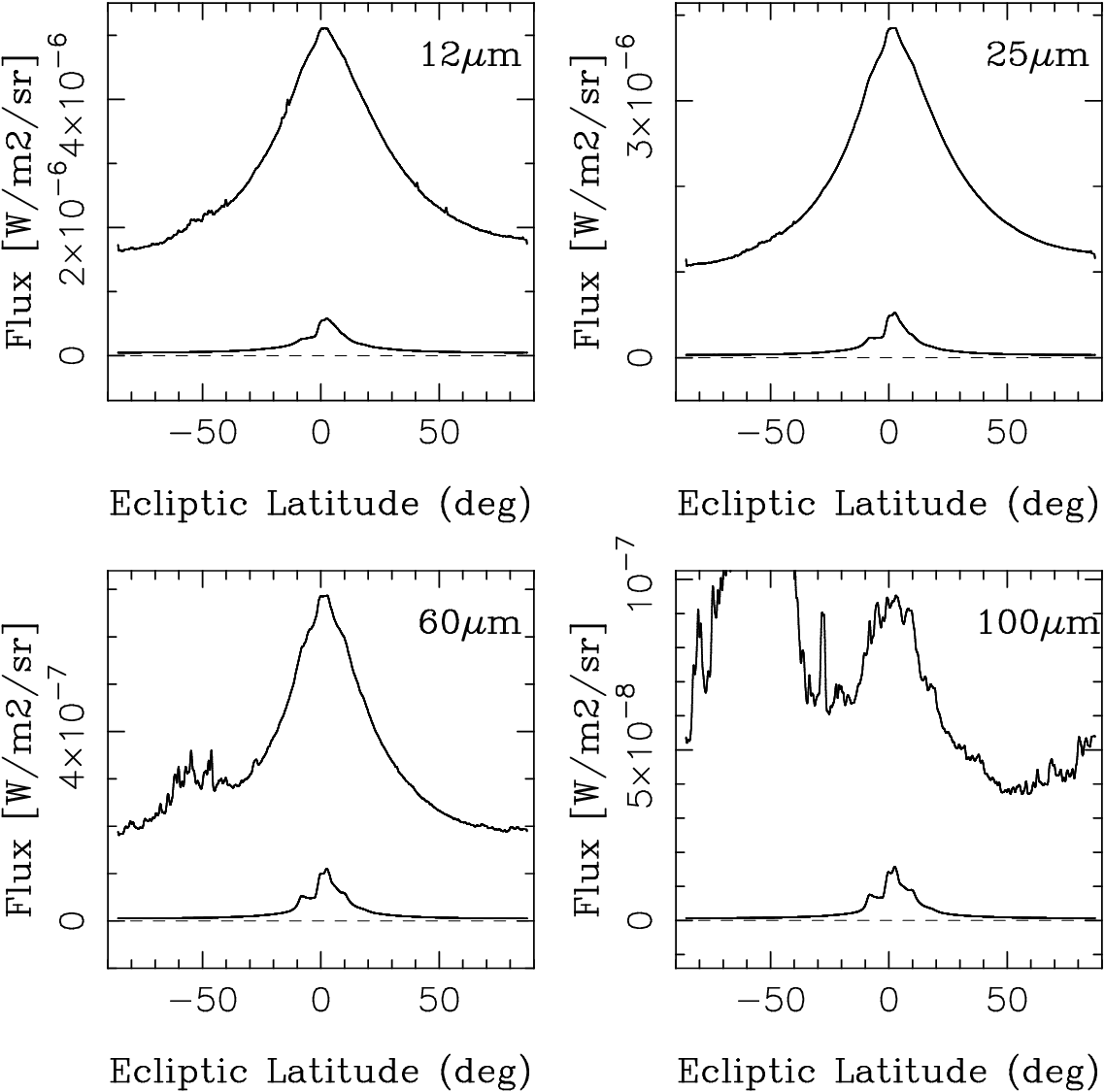}
\caption{The upper solid lines in each panel show IRAS scan 180\_24 (see Table~3 in NVBS06) that 
has been smoothed by a low pass-filter to remove point sources and instrumental noise. 
Different panels show fluxes at 12, 25, 60 and 100 $\mu$m IRAS wavelengths. The bottom solid 
lines show the contribution of three main asteroid dust bands to the observed fluxes. According to
NVBS06, these dust bands contribute to the observed fluxes by $\approx$9-15\% 
within 10$^\circ$ to the ecliptic, and $<$5\% overall. The strong signal at 100 $\mu$m  
between latitudes $b \approx -80^\circ$ and $b \approx -30^\circ$ is the galactic plane 
emission (also apparent at 60 $\mu$m). Figure from NVBS06.}
\label{hats}
\end{figure}

IRAS showed that the MIR brightness of the zodiacal cloud peaks at the 
ecliptic and has broad wings that extend all the way to the ecliptic poles (Fig. \ref{hats}). 
The variation with the ecliptic longitude is minimal indicating that the zodiacal cloud is a 
nearly symmetrical disk of circumsolar particles that is roughly centered at the Sun (e.g.,
Staubach et al., 2001). The significant flux received from the ecliptic poles 
($\sim$1/3-1/4 of the ecliptic flux) also 
suggests that the cloud must be rather thick in the normal direction to the ecliptic plane. 

Following Nesvorn\'y et al. (2006; hereafter NVBS06), we selected several representative 
IRAS scans that met the 
following conditions. (1) We used scans that covered a continuous range of ecliptic latitudes 
from $b<-80^\circ$ to $b>80^\circ$. (2) We did not use 
scans that had gaps created when the telescope skipped over bright sources. (3) We did not 
use scans that showed strong emission from extrasolar sources such as the galactic plane, 
galactic cirrus, point sources, etc. (4) We required that the selected scans covered all 
values of ecliptic longitude and the available range of solar elongations, $l_\odot$. 
Tables~2 and 3 in NVBS06 list the basic information about the selected scans. 

In this work, we only consider scans with $l_\odot \approx 90^\circ$. This is mainly done because 
the detail variation of brightness with $l_\odot$ is difficult to characterize as it requires 
an appropriate model for the collisional disruption of particles (NVBS06). We do not include
the effects of collisions between particles in our model (except in \S4.2, where a simple model 
for collisions is used). The considered value of $l_\odot$ is in the middle of the available 
range and thus best represents IRAS observations.

The zodiacal cloud is known to be warped and have a center that is slightly 
offset from the Sun. These features are produced by gravitational perturbations from
Jupiter (Dermott et al., 1995, 2001). Since we are not interested in these detailed features of 
the zodiacal cloud here, we removed them by combining the selected IRAS scans into a representative 
profile. This was done by first shifting the scans by a small value in latitude ($<$2$^\circ$), so that the 
peak of emission was centered exactly at $b=0$, and calculating the average flux from all 
selected scans as a function of $b$. We ended up with profiles showing the mean fluxes at 12, 25 
and 60 $\mu$m wavelengths as a function of ecliptic latitude (Fig. \ref{mean}). These 
profiles represent the main constraints on the work described here. Additional constraints are
discussed in \S4.3.
\section{Model}
Our model of the zodiacal cloud has four parts: we (i) define the initial orbital distributions
of particles for different sources (asteroid and comet populations); (ii) track the orbital 
evolution of particles with various sizes from sources to sinks; (iii) determine the thermal 
infrared emission from these synthetic particle distributions; and (iv) model the detection of 
their emission by IRAS. These model components are described below.  

To simplify things, we do not initially account for the collisional disruptions of dust particles 
in the model. This is a major assumption which we verify in \S4.2. For example, NVBS06 
showed how the collisional disruption of particles, and production of smaller daughter 
products, affect the spatial distribution of particle populations in the asteroidal dust bands.
The work described here, however, is less sensitive to collisional 
processes because the overall gross shape of the zodiacal cloud should be mainly controlled by 
the orbital distribution of its source population(s), violent dynamics of particles moving on 
planet-crossing orbits, and the effects of PR drag. These features/processes are included 
in our model as we describe below.
\subsection{Sources}
Our model starts with the source population of objects. This may be either of the following: (1) 
individual asteroid groups such as the Karin, Veritas or Beagle asteroid families (NVBS06, 
Nesvorn\'y et al., 2008); (2) asteroid belt as a whole; 
(3) active JFCs defined by their debiased orbital distribution obtained from Levison 
\& Duncan (1997; hereafter LD97) and their physical lifetime; 
(4) JFCs with orbital distribution similar to (3) but dynamically evolved beyond their
nominal active lifespan; (5) Halley-type comets (HTCs); and (6) Oort-cloud comets (OCCs). 
These source populations are described in a more detail below.

We ignore the contribution of interstellar dust particles because thermal emission from 
these small particles (diameter $D<1$ $\mu$m) would create diagnostic spectral features in 
the mid-infrared wavelengths that are not observed (e.g., Reach et al., 2003). More specifically, 
observations of the MIR spectrum of the zodiacal cloud by Reach et al. (2003) suggest that 
the bulk of the zodiacal cloud is produced by $\sim$10-100 $\mu$m particles. We also ignore 
dust produced by disruptive collisions in the Kuiper belt (hereafter KB dust) because Landgraf et 
al. (2002) and Moro-Mart\'{\i}n and Malhotra (2003) showed that KB dust particles should represent 
only a minor contribution to the inner zodiacal cloud. 

Results for (1) were taken from NVBS06 and Nesvorn\'y et al. (2008) who found that the three
main dust bands discovered by Low et al. (1984) originate from the Karin, Veritas and Beagle 
asteroid families. According to these results (Fig. \ref{hats}), the three main dust 
bands represent only $\approx$9-15\% of the zodiacal cloud emission at the ecliptic and 
$<$5\% overall. About a dozen other dust bands have been identified (Sykes, 1990).
Since these dust bands are much fainter than the three main dust bands, Nesvorn\'y et al. results 
set the upper limit on the contribution of {\it identified} asteroid breakups to the zodiacal cloud. 


The orbital distribution of the asteroid belt source (2) is modeled by using the observed 
orbital distribution of asteroids with $D>15$~km. We obtained this distribution from the 
ASTORB catalog (Bowell et al., 1994). This sample should be complete and unbiased 
(Jedicke and Metcalfe, 1998; Gladman et al., 2009). The orbital distribution of asteroids with $D<15$~km, which 
may be a better proxy for the initial distribution of dust produced in main belt collisions 
(Sykes and Greenberg, 1996), is roughly similar to that of large asteroids. Thus, the orbital distribution 
that we use should be a reasonable assumption for (2).
   
The orbital distributions obtained by numerical integrations of test particle trajectories 
from individual comets would suffer from the heavily biased comet catalogs. Moreover, there 
are hundreds of known comets, each producing dust at a variable and typically unknown rate. In 
addition, it is also possible that the present zodiacal dust complex contains particles from 
lost parents as suggested by the orphaned Type-II trails (Sykes, 1990) and identification 
of meteoroid streams with disrupted JFCs (see Jenniskens (2008) for a review). In view of 
these difficulties, we resorted to the following strategy.

The orbital distribution of JFCs was taken from LD97 who followed the evolution of bodies 
originating in the Kuiper belt as they are scattered by planets and evolve in small 
fractions into the inner solar system. Starting at the time when the comets'  perihelion 
distance, $q$, first drops below 2.5 AU in the LD97 simulations, we include it as an active 
JFC in our list of source objects. LD97 showed that the orbital distribution of visible JFCs 
obtained in this way nicely approximates the observed distribution. Moreover, LD97 argued 
that the inclination distribution of new JFCs (reaching $q<2.5$ AU for the first time) is 
relatively narrow. The inclination distribution widens at later times as the JFC orbits 
become more spread by Jupiter encounters. 

By comparing the width of the model inclination distribution with that of the observed JFCs, LD97 
were able to estimate the fading lifetime of JFCs, $t_{\rm JFC}$, defined as the characteristic 
time between their first and last apparitions. They found that $t_{\rm JFC}=12,000$ yr with a 90\% 
confidence interval $3,000<t_{\rm JFC}<30,000$~yr. See Fig. \ref{bin1} for an illustration of the 
steady-state JFC orbit distribution for $t_{\rm JFC}=12,000$ yr. This is our initial
distribution of particles produced by {\it active} JFCs. We use $t_{\rm JFC}$ as a free parameter 
in our model with values extending to $t_{\rm JFC}=100,000$~yr to account for the possibility that 
an important dust component may be produced by old dormant JFCs as they spontaneously disrupt.  

Our models for the orbital distributions of HTCs and OCCs are simpler than the 
one described for JFCs above, because we do not have in hands an appropriate numerical model
that we could use with confidence. Fortunately, this is not a major limitation factor 
in this work because our main results described in \S4 are not sensitive to the detailed 
properties of the HTCs and OCCs populations. 

For HTCs, we assume that the differential distribution of the perihelion distance, ${\rm d}N(q)$,
is ${\rm d} N(q) \propto q\, {\rm d}q$, and set an upper limit of $q$ at 3 AU. HTCs typically become 
visible/active only if they reach $q<3$ AU. Similarly, the cumulative semimajor axis 
distribution of HTCs, $N(<a)$, is assumed to be $N(<a) \propto a$ with an upper cut on $a$ 
at 50 AU. The differential inclination distribution $dN(i)$ is taken from Levison et al. 
(2006). This distribution can be approximately described by
\begin{equation}
{\rm d}N(i) \propto \sin(i) {\rm e}^{-{1\over 2} \left({i\over \sigma}\right)^2}\, {\rm d}i 
\end{equation}
with $\sigma=30^\circ$.

According to Francis (2005), the long-period comets have ${\rm d}N(q) \propto (1 + \sqrt{q})\, 
{\rm d}q$ for $q<2$ AU. For $q > 2$ AU, Francis'  study predicts ${\rm d}N(q)$ being flat or declining
while we would expect the perihelion distribution to increase with $q$. It probably just 
shows that the distribution is not well constrained for $q > 2$ AU. We use ${\rm d}N(q) \propto 
2.41 (q/2)^\gamma\, {\rm d}q$ for $q>2$ AU with $0<\gamma\leq1$. The semimajor axis values of 
OCCs are set between 10,000 and 50,000 AU, which is known as the Oort spike (Wiegert and Tremaine, 
1999). We also use $a=1,000$~AU to check on the dynamical behavior of dust particles launched 
from the returning OCCs. The inclination distribution of OCCs is set to be uniformly random between 
0 and 180$^\circ$.      
 
We use two different methods to launch particles from their source objects. In the first method,
chosen to approximate the ejection of particles from active comets, we launch particles
at the perihelion when the mean anomaly $M=0$. In the second method, we launch
particles along the orbit with uniform distribution in $M$. This second method is more
appropriate for the asteroidal particles and for particles produced by comet disruptions. 
Indeed, the identified comet disruptions do not seem to be correlated in any way with the
perihelion passage (e.g., Weissman, 1980). To simplify things, we neglect the ejection 
velocities of dust particles from their parent object and assume that they will initially 
follow the parent object's orbit modified by the radiation pressure. 

The individual comets in our model are assumed to contribute in roughly the same proportion to the 
circumsolar dust complex. The reasoning behind this assumption is that if an individual super-comet were 
the dominant source of circumsolar dust, the zodiacal cloud would not have such a smooth 
and symmetrical structure. In fact, the observed smooth structure of the zodiacal cloud probably 
implies a source population that contains numerous objects that are well mixed in the orbital 
space. 
\subsection{Orbit Evolution} 
The orbits of particles were tracked using the Wisdom-Holman map (Wisdom and 
Holman, 1991) modified to include effects of radiation forces (Burns et al., 1979; 
Bertotti et al., 2003). The acceleration $\vec{F}$ on a particle due to these forces is 
\begin{equation} \vec{F} = \beta G {m_\odot \over R^2} \left[ \left (1 - {\dot{R} \over 
c} \right) {\vec{R} \over R}  - {\vec{V} \over c} \right] \ , \label{acc} \end{equation}  
where $\vec{R}$ is the orbital radius vector of the particle, $\vec{V}$ is its velocity,  
$G$ is the gravitational constant, $m_\odot$ is the mass of the Sun, $c$ is the speed  of 
light, and $\dot{R} = {\rm d}R/{\rm d}t$. The acceleration (\ref{acc}) consists of the 
radiation pressure and velocity-dependent PR terms. Parameter 
$\beta$ is related to the radiation pressure coefficient, $Q_{\rm pr}$, by 
\begin{equation} \beta = 5.7\times10^{-5}\ {Q_{\rm pr} \over \rho s}\ , \label{beta} \end{equation} 
where radius $s$ and density $\rho$ of the particle are in cgs units. Pressure coefficient 
$Q_{\rm pr}$ can be determined using the Mie theory (Burns et al., 1979). We used 
$Q_{\rm pr} = 1$ which corresponds to the geometrical optics limit where $s$ is much  
larger than the incident-light wavelength. We assumed that the solar-wind drag force has 
the same functional form as the PR term and contributes by 30\% to the total drag intensity 
(Gustafson, 1994).   

We used particles with diameter $D=2s=10$, 30, 100, 200, 300, 1000 $\mu$m and set their bulk density 
to $\rho=2$~g~cm$^{-3}$. For comparison, Love et al. (1994) reported $\rho \approx 2$ g 
cm$^{-3}$ for stratospheric-collected IDPs, while McDonnel and Gardner (1998) found mean 
$\rho = 2$-2.4~g~cm$^{-3}$ from the analysis of data collected by the LDEF and Eureca 
satellites. On the other hand, density of 1 g cm$^{-3}$ has been often assumed for cometary 
matter (e.g., Joswiak et al., 2007; Wiegert et al., 2009). Gr\"un et al. (1985) suggested that 20-40\% of particles may have 
low densities whereas most meteoroids have $\rho=2$-3~g~cm$^{-3}$. Our results may be easily 
rescaled to any $\rho$ value and we explicitly discuss the effect of $\rho$ wherever it is appropriate.

The particle orbits were numerically integrated with the {\tt swift\_rmvs3} code (Levison and 
Duncan, 1994) which is an efficient implementation of the Wisdom-Holman map and which, in 
addition, can deal with close 
encounters between particles and planets. The radiation pressure and drag forces were 
inserted into the Keplerian and kick parts of the integrator, respectively. The 
change to the Keplerian part was trivially done by substituting  $m_\odot$ by 
$m_\odot(1-\beta)$, where $\beta$ is given by Eq. (\ref{beta}). 

The code tracks the orbital evolution of a particle that revolves around the Sun and is 
subject to the gravitational perturbations of seven planets (Venus to Neptune) until the 
particle impacts a planet, is ejected from the solar system or drifts to within 0.03~AU 
from the Sun. We removed particles that evolved to $R<0.03$~AU because the orbital period 
for $R\lesssim0.03$~AU is not properly resolved by our integration timestep. In 4.2, we also 
consider the effect of collisional disruption by removing particles from the simulations 
when they reach their assumed physical lifespan. We followed 1,000-5,000 particles for each 
$D$, source, and parameter value(s) that define that source.
\subsection{Thermal Emission of Particles} 
Particles were assumed to be isothermal, rapidly rotating spheres. The absorption was 
assumed to occur into an effective cross-section $\pi s^2$, and emission out of $4 \pi 
s^2$. The infrared flux density (per wavelength interval ${\rm d} \lambda$) per unit surface 
area at distance $r$ from a thermally radiating particle with radius $s$ is 
\begin{equation} F(\lambda) = \epsilon(\lambda,s) B(\lambda,T) {s^2 \over r^2} \ , 
\label{flux} \end{equation} 
where  
$\epsilon$ is the emissivity and $B(\lambda,T)$ is the energy flux  at 
$(\lambda,\lambda+{\rm d}\lambda)$ per surface area from a black body at temperature 
$T$: 
\begin{equation} B(\lambda,T) = {2 \pi h c^2 \over \lambda^5} \left[ {\rm 
e}^{hc/\lambda k T} - 1 \right]^{-1} \ . \label{bb} \end{equation} 
In this equation, $h=6.6262\times10^{-34}$ J s is the Planck constant, $c = 2.99792458 
\times 10^8$ m s$^{-1}$ is the speed of light, and $k = 1.3807\times10^{-23}$ J K$^{-
1}$ is the Boltzmann constant. We used $\epsilon=1$ which should be roughly appropriate
for the large particles used in this work. See NVBS06 for a more precise 
treatment of $\epsilon(\lambda,s)$ for dust grains composed of different materials.

To determine the temperature of a particle at distance $R$ from the Sun, we used the 
temperature variations with $R$ that were proposed by different authors from spectral 
observations of the zodiacal cloud (e.g., Dumont et al., 1988; Renard et al., 1995; 
Leinert et al., 2002; Reach et al., 2003). For example, Leinert et al. (2002) proposed 
that $T(R) = 280/R^{0.36}$~K near $R=1$~AU from ISOPHOT spectra. We 
used $T(R) = T_{\rm 1 AU}/R^\delta$ K, where $T_{\rm 1 AU}$ is the temperature 
at 1~AU and $\delta$ is a power index. We varied $T_{\rm 1 AU}$ and $\delta$ to see 
how our results depend on these parameters. Values of $T_{\rm 1 AU} \approx 280$~K and $\delta=0.5$ 
correspond to the equilibrium temperature of large dark particles. Values $\delta<0.5$ would 
be expected, for example, for fluffy particles with small packing factors (e.g., Gustafson et al., 2001).
\subsection{Synthetic Observations} 
To compare our results with IRAS observations described in \S2, we developed a code that 
models thermal emission from distributions of orbitally evolving particles and produces infrared
fluxes that a space-borne telescope would detect depending on its location, pointing 
direction and wavelength. See NVBS06 for a detailed description of the code. 

In brief, we define the brightness integral along the line of sight of an infrared 
telescope (defined by fixed longitude $l$ and latitude $b$ of the pointing 
direction) as:
\begin{equation}
\int_{a,e,i} {\rm d}a {\rm d}e {\rm d}i
\int_{0}^{\infty} {\rm d}r\ r^2 
\int_D {\rm d}D\, F(D,r) N(D;a,e,i) S(R,L,B) \ ,
\label{all}
\end{equation}
where $r$ is the distance from the telescope, $F(D,r)$ is the infrared flux 
(integrated over the wavelength range of the telescope's system) per unit surface 
area at distance $r$ from a thermally radiating particle with diameter $D$. 
$S(R,L,B)$ defines the spatial density of particles as a function of the
heliocentric distance, $R$, longitude, $L$, and latitude, $B$ (all functions 
of $r$ as determined by geometry from the location and pointing direction 
of the telescope). $N(D;a,e,i)$ is the number of particles having effective 
diameter $D$ and orbits with semimajor axis, $a$, eccentricity, $e$, and 
inclination, $i$. 

We evaluate the integral in Eq. (\ref{all}) by numerical renormalization (see NVBS06). $F(D,r)$ is 
calculated as described in \S3.3. $N(D;a,e,i)$ is obtained from numerical simulations 
in \S3.2. $S(R,L,B)$ uses theoretical expressions for the spatial distribution of particles with 
fixed $a$, $e$ and $i$, and randomized orbital longitudes (Kessler, 1981; NVBS06). 

We assume that the telescope is located at $(x_t=r_{\rm t} \cos \phi_{\rm t},y_t=r_{\rm t} 
\sin \phi_{\rm t},z_{\rm t}=0)$ 
in the Sun-centered reference frame with $r_{\rm t} = 1$ AU. Its viewing direction is defined by a 
unit vector with components $(x_{\rm v},y_{\rm v},z_{\rm v})$. In Eq. (\ref{all}), the pointing vector can be also 
conveniently defined by longitude $l$ and latitude $b$ of the pointing direction, where $x_{\rm v} = \cos b \cos 
l$, $y_{\rm v} = \cos b \sin l$, and $z_{\rm v} = \sin b$. As described in \S2, we fix the solar elongation 
$l_\odot = 90^\circ$ and calculate the thermal flux of various particle populations as a 
function of $b$ and wavelength. The model brightness profiles at 12, 25 and 60 $\mu$m are 
then compared with the mean IRAS profiles shown in Fig.~\ref{mean}. 

To check our code, known as Synthetic InfraRed Telescope (SIRT), we programmed a particle 
version of the algorithm, which should be in many ways similar to the core algorithm in SIMUL 
(Dermott et al. 1988; see also Dermott et al., 2001). The particle version inputs the orbital elements of particles 
obtained in the orbital simulations (\S3.2) and produces their orbit clones that are uniformly 
spread over $2\pi$ in mean anomaly $M$. Thus, every test particle is assumed to trace a cloud of 
real particles having the same orbit as the test particle but different angular locations along 
the orbit. See Vokrouhlick\'y et al. (2008; their \S2.5) for a technical description of the algorithm. 
This procedure is based on the assumption that any concentration of particles with a specific $M$ value
would be quickly dispersed by the Keplerian shear. We employ this procedure to improve the resolution. 
Without it, the number of integrated test particles would be too small to obtain a useful result.

To be able to compare the results of the particle algorithm with SIRT, the particle algorithm must
also use smooth distributions in perihelion longitude $\varpi$ and nodal longitude $\Omega$. This 
is achieved by generating additional orbital clones with $\varpi$ and $\Omega$ uniformly spread
over $2\pi$. Figure \ref{check} shows examples of the results obtained from the particle 
algorithm and SIRT. The agreement between the two codes is excellent which gives us confidence
that both codes work properly. We find that the SIRT algorithm based on the Kessler distribution 
is much faster than the particle one. For this reason, we use the original SIRT code in this study. 
\section{Results}
Our primary model parameters are the relative contribution of different sources to the zodiacal 
cloud. The total model flux as a function of the ecliptic latitude is 
obtained as
\begin{equation}
F_{\rm model}(b) = \alpha_{\rm AST} F_{\rm AST} + \alpha_{\rm JFC} F_{\rm JFC} + \alpha_{\rm HTC} 
F_{\rm HTC} + \alpha_{\rm OCC} F_{\rm OCC}\ ,
\label{fm}
\end{equation}
where $\alpha$ are coefficients that satisfy $\alpha_{\rm AST}+\alpha_{\rm JFC}+\alpha_{\rm HTC}+
\alpha_{\rm OCC}=1$, and $F_{\rm AST}$, $F_{\rm JFC}$, $F_{\rm HTC}$ and $F_{\rm OCC}$ are model fluxes obtained 
for different sources. We normalize them so that the ecliptic model flux from each source
is equal to that of the mean observed flux at $b=0$. Coefficients $\alpha$ therefore give
the relative contribution of different sources at the ecliptic.  

The model flux profiles depend on the particle size, wavelength, and for JFCs also 
on the assumed value of $t_{\rm JFC}$. As described in \S3.2, we tracked particles with $10<D<1000$~$\mu$m. 
These different sizes are treated individually in Eq. (\ref{fm}). In particular, we do not attempt to 
construct plausible size-frequency distributions for different sources. It is therefore assumed  
that a single characteristic particle size, or a narrow range of sizes, can be effectively used to model
the observed MIR flux. This assumption needs to be verified later.

In 4.1, we first consider a model where the lifespan of particles is limited by their dynamical lifetime.
Effects of particle disruptions are discussed in 4.2. 
\subsection{Collision-Free Model}
Figure \ref{mod1} shows the 25-$\mu$m flux of $D=100$ $\mu$m particles produced by different 
source populations. Note that these profiles do not sensitively depend on the particle size 
(see Figs. \ref{ast1} and \ref{dep} for the results for different $D$). Instead, the main differences between 
results for different source populations in Fig. \ref{mod1} reflect the initial orbit distribution of particles in each 
source and their orbit evolution. Therefore, these profiles can help us to identify the source population(s) 
that can best explain IRAS observations.

The asteroidal particles produce a profile with a very sharp peak centered at the ecliptic.
The emission from asteroidal particles near the ecliptic poles is relatively faint. The polar emission comes 
from the particles that evolved by PR drag from $a>2$ AU to $R=1$ AU. While most asteroidal particles 
indeed reach 1 AU, they pass too briefly near $b=\pm90^\circ$ to produce important polar fluxes. 
This is why most radiation is received from $b\sim0$, where the telescope collects the thermal
emission of particles over a wide distance range. A broader distribution of orbital inclinations 
is apparently needed to match IRAS measurements.
 
The profile produced by HTC particles is much broader than the observed one (Fig. \ref{mod1}). 
In this case, the magnitude 
of the polar fluxes is $\approx$1/2 of that near the ecliptic. This result reflects the very broad 
inclination distribution of HTCs (\S3.1). A potentially significant contribution of HTCs 
to the zodiacal cloud is also problematic because the two large HTCs, 109P/Swift-Tuttle and 1P/Halley, 
tend to librate about mean-motion resonances, causing relatively stable orbits for long periods 
of time. Thus, dust released by HTCs is expected to be concentrated along certain location on the 
sky making it difficult to explain the smooth profile of the zodiacal dust. Note also that 
Altobelli et al. (2007) have not detected HTC particle impacts in the Cassini dust experiment,
indicating that HTC particles are relatively sparse.

The OCC particles, which have a nearly isotropic inclination distribution, produce the MIR flux 
that is constant in latitude (not shown in Fig. \ref{mod1}). Therefore, the ecliptic and polar fluxes from 
OCC particles are roughly the same and do not match observations. We conclude that a single-source model with 
either asteroidal, HTC or OCC particles cannot match the observed profile of the zodiacal cloud. 

We are left with JFCs. It is notable that the width and shape of the JFC profile in Fig. \ref{mod1} 
closely matches observations. The only slight difference is apparent for large ecliptic latitudes
where the model flux, shown here for $D=100$ $\mu$m and $t_{\rm JFC} = 12,000$ yr, is slightly 
weaker than the one measured by IRAS. We will discuss this small difference below and show that it 
could be explained if: (1) slightly smaller JFC particles were used, and/or (2) the zodiacal cloud has 
a faint isotropic component. We thus believe that the close resemblance of our model JFC profile 
with IRAS data is a strong indication that {\it JFCs are the dominant source of particles in the 
zodiacal cloud.} 

Since asteroids and active JFCs have similar inclination distributions (Hahn et al., 2002), it 
may seem surprising that JFC particles produce substantially wider MIR flux profiles than asteroidal 
particles. By analyzing the results of our numerical integrations we find that the encounters with 
terrestrial planets and secular resonances are apparently not powerful enough to significantly affect 
the inclination distribution of drifting asteroidal particles. The inclination distributions of 
the asteroidal particles and their source main-belt asteroids are therefore essentially the same
($\approx$10$^\circ$ mean $i$). On the other hand, we find that JFC particles are scattered by Jupiter 
before they are able to orbitally decouple from the planet and drift down to 1 AU. This results in a 
situation where the inclination distribution of JFC particles is significantly broader 
($\approx$20$^\circ$ mean $i$ for $R<3$ AU) than that of their source JFCs. This explains Fig. 
\ref{mod1} and shows limitations of the arguments about the zodiacal cloud origin based on the 
comparative analysis of sources (e.g., Hahn et al., 2002).

We will now address the question of how the MIR fluxes from the JFC particles depend on $D$ and 
$t_{\rm JFC}$. We define:
\begin{equation}
\eta^2 = {1 \over \pi} \int_{-\pi/2}^{\pi/2} {\left[{F_{\rm IRAS}(b)-F_{\rm model}(b)}\right]^2 \over 
\sigma^2(b)}\ {\rm d}b\ ,
\label{eta}
\end{equation}
where $F_{\rm IRAS}$ is the mean IRAS flux, $\sigma$ is the standard deviation of $F_{\rm IRAS}$ determined 
from the spread of representative IRAS scans for each $b$ (\S2), and $F_{\rm model}=F_{\rm JFC}$ (i.e., 
$\alpha_{\rm JFC}=1$ and $\alpha_{\rm AST}=\alpha_{\rm HTC}=\alpha_{\rm OCC}=0$ in Eq. (\ref{fm})). Note that the 
integration in Eq. (\ref{eta}) is set to avoid the intervals in $b$ with strong galactic 
emission. 

While the definition of $\eta^2$ in Eq. (\ref{eta}) is similar to the usual $\chi^2$ statistic (e.g., 
Press et al., 1992), we will not assign a rigorous probabilistic meaning to the $\eta^2$ values  
obtained from Eq. (\ref{eta}). This is mainly because it is not clear whether the $\sigma$ values computed 
in \S2 from the IRAS data can adequately represent the measurement errors. Instead, we will use Eq. (\ref{eta}) 
only as an indication of whether a particular model is more reasonable than another one. Models with 
$\eta^2 \lesssim 1$ will be given priority. For a reference, the JFC model in Fig. \ref{mod1} gives 
$\eta^2 = 5.1$.  

We calculated $\eta^2$ as a function of $D$ and $t_{\rm JFC}$ to determine which values of these 
parameters fit IRAS observations best. We found that the best fits with $\eta^2 < 10$ occur for 
$30\leq D \leq100$ $\mu$m and $t_{\rm JFC} \leq 30,000$ yr. 

Figure \ref{dep} illustrates how the shape of the 25-$\mu$m profile produced by JFC particles depends 
on $D$ and $t_{\rm JFC}$. The profiles become wider with increasing $D$ and $t_{\rm JFC}$ values.
For $t_{\rm JFC}=12,000$ yr, the best results were obtained with $D=30$ $\mu$m and $D=100$ $\mu$m 
($\eta^2 = 3.2$ and 5.1, respectively). The profiles for $D=10$ $\mu$m are too narrow and clearly do 
not fit data well ($\eta^2 = 70$), while those for $D=1000$ $\mu$m are slightly too wide ($\eta^2 = 58$). 
We also found that there are no really good fits with $t_{\rm JFC} > 30,000$ yr, because the profiles 
are too broad near the ecliptic independently of the particle size. 

The best single-source fits discussed above have $\eta^2 > 1$ which is not ideal. According to our 
additional tests this is probably not due to the coarse resolution and studied range of $D$ and 
$t_{\rm JFC}$. Instead, this may point to: (1) a subtle problem with our JFC model, and/or (2) 
the possibility that additional minor sources (such as asteroids, OCCs or HTCs) should be included in the 
model. Option (1) is difficult to test unless a better model of the JFC population becomes available.
Here we concentrate on (2) because an ample evidence exists (e.g., for $\sim$10\% near-ecliptic 
asteroid contribution from asteroid dust band modeling) that these additional sources may be relatively 
important.

We start by discussing the results obtained by assuming that the zodiacal cloud has two sources. The 
motivation for considering the two-source model was the following. First, we wanted see whether a 
combination of two sources could successfully fit the observed profile. Second, we attempted to 
place upper limits on the contributions of asteroid, OCC and HTC sources. While it is obvious that models 
with more than two sources can be tuned to fit the data better, it is not clear whether more than 
two sources are actually required. Our two-source models were used to test these issues. 

In the first test, we used the two-source model with asteroids and OCCs (i.e., $\alpha_{\rm JFC}=\alpha_{\rm HTC}=0$
in Eq. (\ref{fm})). We found that this particular model produces unsatisfactory fits ($\eta^2>10$) to IRAS 
observations for all particle sizes considered here (10-1000 $\mu$m) (Fig. \ref{fit2}a). The model profile 
is significantly narrower near the ecliptic, where the asteroid component prevails, and is too wide for 
$|b|\gtrsim40^\circ$, where the OCC component prevails. This happens mainly due to the fact that the 
asteroid dust is confined to the ecliptic plane and produces a very narrow profile near the ecliptic
(Fig. \ref{mod1}). We also find it unlikely that two so distinct populations of objects, such as the main 
belt asteroids and OCCs, would have comparable dust production rates. Thus, we believe that the 
two-source model with asteroid and OCC dust can be dismissed.

In the second test, we set $\alpha_{\rm OCC}=\alpha_{\rm HTC}=0$ and considered models of the zodiacal cloud with 
the JFC and asteroid components. We found that a small contribution of asteroid dust can improve the
fits. For example, $\eta^2=0.92$ for the $D=30$-$\mu$m JFC particles with $\alpha_{\rm JFC}=
0.9$ and $t_{\rm JFC}=12,000$ yr, and $D=100$-$\mu$m asteroidal particles with $\alpha_{\rm AST}= 0.1$. This 
represents a significant improvement from $\eta^2=3.2$ that we obtained for a single-source model with JFC 
particles only. Values $\alpha_{\rm AST} \gtrsim 0.3$ are clearly unsatisfactory because $\eta^2>10$ for 
$\alpha_{\rm AST}>0.3$. Also, $\eta^2>3.1$ for $\alpha_{\rm AST}>0.2$ which shows that the fit does not improve 
when we add a $\gtrsim$20\% asteroid contribution. These results suggests that a very large asteroid contribution 
to the zodiacal cloud can be ruled out. This agrees with the conclusions of NVBS06 who found that 
$\alpha_{\rm AST} \lesssim 0.15$ from modeling of the main asteroid dust bands.

Finally, the two-component model with the JFC and isotropic OCC sources can fit data very well 
(Fig. \ref{fit2}b). With $D=100$ $\mu$m and $t_{\rm JFC} = 12,000$ yr, corresponding to one of 
our best single-source fits with JFCs, $\alpha_{\rm JFC}=0.97$ and $\alpha_{\rm OCC}= 0.03$, we find 
that $\eta^2 = 0.36$, by far the best fit obtained with any two-source model. As Fig. \ref{fit2}b
shows, the fit is excellent. We may thus find an evidence for a small contribution 
of OCC particles to the zodiacal cloud. A much larger OCC contribution is not supported by the data 
because the fit gets significantly worse for $\alpha_{\rm OCC}>0.1$. For example, $\eta^2 > 10$ for 
$\alpha_{\rm OCC}>0.15$ which is clearly unsatisfactory. A large contribution of OCC particles 
can therefore be rejected.

We propose based on the results described above that the zodiacal cloud has a large JFC component
($\alpha_{\rm JFC}\gtrsim0.9$), and small asteroid/OCC components ($\alpha_{\rm AST} \lesssim 0.2$ and 
$\alpha_{\rm OCC} \lesssim 0.1$). To verify this conclusion, we considered three-component models with $\alpha_{\rm HTC}=0$
and used $\alpha_{\rm JFC}$, $\alpha_{\rm AST}$ and $\alpha_{\rm OCC}$ as free parameters. We found that the best 
two-source fit with $\alpha_{\rm JFC}=0.97$ and $\alpha_{\rm OCC}= 0.03$ cannot be significantly improved by including a 
small asteroid contribution. Similarly, the fit with $\alpha_{\rm JFC}=0.9$ and $\alpha_{\rm AST}=0.1$ cannot
be improved by adding a small OCC contribution. Thus, the asteroid/OCC contributions cannot be constrained
independently because their effects on the combined profiles can be compensated by adjusting 
the $D$ and $t_{\rm JFC}$ values of the dominant JFC particles.

If we set the parameters of the dominant JFC particles to be $D=100$ $\mu$m and $t_{\rm JFC} = 12,000$ yr,
however, the $\alpha_{\rm AST}$ and $\alpha_{\rm OCC}$ values can be constrained much better (Fig. \ref{alphas}).
For example, models with $\eta^2<1$ require that $\alpha_{\rm AST}<0.22$ and $\alpha_{\rm OCC}<0.13$. Thus,
under reasonable assumptions, the contribution of asteroid particles to the near-ecliptic IRAS fluxes is
probably $<10$-20\%, in agreement with the results obtained in NVBS06 from modeling of the asteroid
dust bands. This means that asteroid dust contributes only by $<$10\% to the overall zodiacal dust 
emission at MIR wavelengths. The thermal emission of OCC particles can constitute as much as $\sim$10\% 
of the near-ecliptic emission with $\approx$5\% providing the best fits (Fig. \ref{alphas}). When integrated 
over latitude, the overall OCC component in the zodiacal cloud is likely to be $\lesssim$10\%.

For the sake of consistency with the results suggested from modeling of the asteroid dust bands
(see \S3.1; NVBS06), we impose a small asteroid contribution in the JFC/OCC model. Figure~\ref{fit4} 
shows our preferred fit at different IRAS wavelengths. The $\eta^2$ values of this fit 
in different wavelengths are the following: 0.29 at 12 $\mu$m, 0.35 at 25 $\mu$m and 0.06 at 60~$\mu$m.
This is very satisfactory. Since our model does not include detailed emissivity properties of dust grains 
at different wavelengths (\S3.3), we set the emissivity at 25 $\mu$m to be 1 and fit for the 
emissivities at 12 and 60 $\mu$m. We found that the relative emissivities at 12 and 60 $\mu$m 
that match the data best are 0.76 and 0.87, respectively. Such a variability of MIR emissivity values 
at different wavelengths is expected for $D\approx100$~$\mu$m silicate particles with some carbon content 
(NVBS06). Note also that our preferred $D$ values ($D\approx100$ $\mu$m) are within the range of dominant 
sizes of particles at 1 AU as determined from spacecraft impact experiments ($D=20$-200 $\mu$m; 
Gr\"un et al., 1985).
\subsection{Effect of Disruptive Collisions}
The observational evidence for collisional disruption of interplanetary particles is undeniable (see, e.g., 
Gr\"un et al., 1985), yet it is very difficult to model the full collisional cascade in a computer code as 
each disrupted dust grain produces numerous fragments. The exponentially increasing number of particle 
fragments, which in reality exceeds $10^{25}$ for $D>30$~$\mu$m (NVBS06), renders any full $N$-body integration
impossible. To circumvent this problem, the $N$-body integration of a smaller number of ``tracer'' particles can 
be coupled with a Monte-Carlo model for collisions as in NVBS06. This method is not ideal. Also, any model for 
the collisional cascade would suffer from our lack of detailed understanding of particle properties and 
their fragmentation during impacts. 

Here we opt for a very simple approximation of the effect of disruptive collisions. We assume that the 
collisional lifetime of particles is $t_{\rm col}(D)$ and stop the $N$-body integration of diameter $D$ 
particles when $t=t_{\rm col}(D)$. Thus, particles keep the same $D$ for $t<t_{\rm col}(D)$ and vanish at 
$t=t_{\rm col}(D)$. This is very a crude approximation of the real collisional cascade, in which particles can be 
eroded by small collisions and do not vanish upon disruptive impacts (but produce a range of new particle 
sizes). Also, $t_{\rm col}(D)$ should be a function of $R$ while we assume here that it is not. Still, as 
we show below, our simple model should be able to capture the main effects of particle collisions.

Our choice of $t_{\rm col}(D)$ is motivated by the published estimates of the collisional lifetime of particles 
based on satellite impact rates and meteor observations. For example, Gr\"un et al. (1985) argued that the 
collisional lifetime of $D=1$ mm particles at 2.5 AU should be $\sim$10$^4$ yr (see also Jacchia and 
Whipple, 1961). This relatively short lifetime is a consequence of the dominant population of 
$D\approx100$-300 $\mu$m particles in the inner solar system (e.g., Love and Brownlee, 1993) that are 
capable of disrupting mm-sized particles upon impacts. 

For comparison, the approximate PR drag timescale of particles to spiral down from 2.5 AU to 1 AU is $t_{\rm PR} 
= 2500\ {\rm yr}\times\rho s$, which for $\rho=2$ g cm$^{-3}$ and $s=500$ $\mu$m gives $t_{\rm PR}=2.5\times10^6$ yr. 
Thus, the PR drag lifetime of these large particles is significantly longer than $t_{\rm col}$, indicating 
that they must disrupt before they can significantly evolve by PR drag. Using this assumption in the model
we found that the profiles produced by large JFC particles with $t_{\rm col}\sim10^4$ yr are much narrower 
in latitude than the ones we obtained in \S4.1. This is because large particles die before they can evolve to 
$R\sim1$ AU, where they could contribute to polar fluxes. The zodiacal cloud cross-section therefore cannot be 
dominated by large JFC grains. The large grains are important to explain radar and optical observations of 
meteors (see \S5.3; Taylor and Elford, 1998; Jenniskens, 2006, Wiegert et al., 2009).  

On the other side of the size spectrum, $D<10$ $\mu$m particles have $t_{\rm PR} \ll t_{\rm col}$ due to the 
lack of small $D<1$ $\mu$m impactor particles that are blown out of the solar system by radiation pressure, and because 
the PR drag timescale is short for small $D$ (see, e.g., Dermott et al., 2001). Thus, the small particles are expected to 
spiral down by PR drag from their initial orbits to $R < 1$ AU without being disrupted.
Our original results described in \S4.1 are therefore correct for small particles. We showed in \S4.1 that $D<30$ 
$\mu$m JFC particles do not fit IRAS observations well.

Since $t_{\rm PR} \ll t_{\rm col}$ for small particles and $t_{\rm PR} \gg t_{\rm col}$ for large ones, there 
must exist an intermediate particle size for which $t_{\rm PR} \sim t_{\rm col}$. These intermediate-size particles 
are expected to be very abundant in the zodiacal cloud simply because they have the longest lifetimes. Gr\"un et al. 
(1985) and others argued that the transition from the PR-drag to collision-dominated regimes must happen near 
100 $\mu$m. This is consistent with the LDEF measurements which imply that the $D\approx200$-$\mu$m particles
represent the dominant mass fraction at $R=1$ AU (Love and Brownlee, 1993).

The question is therefore how to model collisional effects for $D\sim100$ $\mu$m. This is not a simple problem 
because the effects of the full collisional cascade, including gradual erosion of particles and their supply 
from breakups of the large ones, should be particularly important in this transition regime. It would be incorrect, 
for example, to take the Gr\"un et al. estimates of $t_{\rm col}$ at their face value and remove particles when 
$t>t_{\rm col}$. In reality, each particle can accumulate PR drift during previous stages of evolution when it is 
still attached to its (slightly) larger precursor particles. 

To test these issues, we assumed a wide range of effective $t_{\rm col}$ and calculated model JFC profiles for
each of these cases. Fig. \ref{tcol} shows that the profiles obtained with $t_{\rm col} \leq 5\times10^5$~yr 
are significantly narrower than the IRAS profiles, even if we tried to compensate for part of the apparent 
discrepancy by OCC particles (Fig. \ref{tcol}b). On the other hand, profiles with $t_{\rm col} \geq 6\times10^5$ yr 
are almost indistinguishable from the original results that we obtained in \S4.1 with $t_{\rm col}=\infty$. 
The transition between $5\times10^5$ yr and $6\times10^5$ yr occurs near the mean PR-drag lifetime of 
$D=100$ $\mu$m JFC particles in our model. It clearly makes a larger difference whether particles are allowed 
to drift down to $R=1$ AU or not. The main lesson we learn from this exercise is that {\it IRAS observations 
imply that the zodiacal cloud particles have been significantly affected by PR drag}.
\subsection{Additional Constraints} 
Additional constraints on the micrometeoroid environment near 1 AU are provided by radar and optical observations 
of meteors. For example, Hunt et al. (2004) determined the meteor entry speeds from the high-gain ALTAIR radar. 
For 30 km s$^{-1}$, the minimum detectable mass is 10$^{-6}$ g (corresponding to $D=100$ $\mu$m for $\rho=2$ g cm$^{-3}$), 
while only mm-sized and larger meteoroids can be detected by the ALTAIR radar for $<$20 km s$^{-1}$. The ALTAIR
measurements represent a significant improvement in sensitivity relative to that of previous radar programs 
(e.g., Taylor, 1995; Taylor and Elford, 1998). For example, Taylor (1995) cited the minimum detectable mass  
of 10$^{-4}$ g at 30 km s$^{-1}$ for the Harvard meteor radar, corresponding to $D\gtrsim500$ $\mu$m particles. 

In \S5.3, we estimate that the mean atmospheric entry speed of $D\sim100$ $\mu$m zodiacal cloud particles is 
$\approx$14 km s$^{-1}$, and that $>$90\% impact at $<$20 km s$^{-1}$. Thus, the ALTAIR and Harvard radars 
cannot detect the bulk of small zodiacal cloud particles impacting Earth at low speeds. 
These measurements are instead sensitive to large meteoroids, which carry
relatively little total mass and cross-section, have short $t_{\rm col}$, and are expected to impact on JFC-like 
orbits. This explains why radar observation show little evidence for populations of small particles with orbits strongly 
affected by PR drag. In Fig. \ref{radar}, we compare the atmospheric entry speeds of $D=1$-mm JFC particles with 
$t_{\rm col}=10^{4}$ yr with the Harvard radar data. This figure documents the dominant role of large JFC 
particles in meteor radar observations.

The spatial distribution of sporadic meteors shows several concentrations on the sky known as the helion/anti-helion,
north/south apex and north/south toroidal sources (e.g., Younger et al., 2009, and the references therein). 
Wiegert et al. (2009) have developed a dynamical model to explain these observations. Their main result concerns
the prominent helion/anti-helion sources for which the large JFC particles clearly provide the best match. Our model for 
large JFC particles is in many ways similar to that of Wiegert et al. (2009). It should therefore be consistent with 
the observed relative strength of the helion/anti-helion sources. The more recent high-gain antenna observations show 
that smaller meteoroids appear to show a weaker helion/anti-helion source of eccentric short-period orbits (Mathews et al., 
2001; Hunt et al., 2004; Galligan and Baggaley, 2004). This implies that orbits of smaller particles should be more  
affected by PR drag (in agreement with \S4.2).

The motion of interplanetary particles can be probed by high-resolution spectral observations of the zodiacal cloud.
Reynolds et al. (2004) measured the profile of the scattered solar Mg I $\lambda$5184 Fraunhofer line in the zodiacal
cloud. The measurements indicate a significant population of dust on eccentric and/or inclined orbits. In particular,
the inferred inclination distribution is broad extending up to about 30$^\circ$-40$^\circ$ with respect to the ecliptic 
plane. The absence of pronounced asymmetries in the shape of the profiles limits the retrograde population of 
particles to less than 10\% of the prograde population. 

These results are in a broad agreement with our model. As we discussed in \S4.1, small JFC 
particles are scattered by Jupiter before they are able to orbitally decouple from the planet and drift down to 
1 AU. This results in a situation where the inclination distribution of JFC particles is broad and extends beyond 
20$^\circ$. The model eccentricities of JFC particles show a broad range of values with most having 
$e=0.1$-0.5 (see \S5.3). This is in a nice agreement with the analysis of Ipatov et al. (2008) who found that 
$e\sim0.3$ best fits the Reynolds et al. data. 

%
\section{Implications}
Given the results described in \S4 we are now in the position to estimate the total cross-section and 
mass of particles in the zodiacal cloud, the current and historical accretion rates of dust by planets
and the Moon, and discuss the implications of our work for studies of micrometeorites and debris disks. 
We address these issues below. 
\subsection{Zodiacal Cloud Mass}
According to our preferred model with the dominant contribution of JFC particles to the zodiacal cloud,
the inner circumsolar dust complex has the total cross-section area of $(2.0 \pm 0.5)\times10^{11}$ km$^2$. 
This is a factor of $\sim$10 larger than the cross-section of asteroidal particles in the main asteroid dust 
bands (NVBS06). The uncertainty of the total cross-section was estimated from the range of values obtained 
for models with $\eta^2<3$. Also about 40\% of the total estimated cross section of the zodiacal cloud, or 
$\approx$$8\times10^{10}$ km$^2$, is in particles that reside inside Jupiter's orbit (i.e., with $R<5$ AU). 

The estimated values are comparable to the effective emitting area of the zodiacal 
cloud defined as 1 ZODY in Gaidos (1999) (1 ZODY $= 1.7\times10^{11}$ km$^2$, assuming blackbody emission 
at 260 K and a bolometric luminosity of $8\times10^{-8}$ $L_\odot$, where $L_\odot$ is the Sun's value; 
Reach et al., 1996; Good et al., 1986). Note, however, that we estimate in \S5.5 that the bolometric 
luminosity of the inner zodiacal cloud is $\sim$2.5 times larger than the one assumed by Gaidos (1999). 

The total mass of the zodiacal cloud is a function of the unknown particle density and loosely constrained 
dominant particle size. With $\rho=2$ g cm$^{-3}$ and $D=200$ $\mu$m, we estimate that the total mass is 
$m_{\rm ZODY} = 5.2\times10^{19}$ g, which is roughly equivalent to that of a 37-km-diameter body. The zodiacal 
cloud therefore currently contains relatively little mass. Note that these estimates apply to the inner
part of the circumsolar dust complex that is detectable at MIR wavelengths. The outer circumsolar dust complex 
beyond Jupiter is likely more massive due to the contribution from KB particles (e.g., Landgraf et al., 
2002; Moro-Mart\'{\i}n and Malhotra, 2003). According to Greaves et al. (2004), the KB dust disk may represent 
up to $\sim$$1.2\times10^{23}$ g. This is up to $\sim$$3\times10^3$ times the mass of the inner zodiacal 
cloud estimated here. Note that this is an upper bound only and that the real KB dust disk can be much less
massive. 


Our mass estimate is at least a factor of $\sim$2 uncertain. For example, if $\rho=1$ g cm$^{-3}$ or 
$D=100$ $\mu$m, we find that $m_{\rm ZODY} = 2.6\times10^{19}$ g. These values are a factor of $\sim$2-4 
lower than the mass of the zodiacal cloud suggested by NVBS06 from modeling of the asteroid dust bands. NVBS06 
assumed that the radial distribution of zodiacal particles is similar to that of the asteroid dust bands, 
which is incorrect if JFCs are the dominant source. On the other hand, NVBS06 determined the realistic size 
distribution of zodiacal particles by tracking the collisional evolution, while we used the single-size 
distributions here.  

We estimate that $\gtrsim$80\% of the total mass at $<$5 AU should be contained in JFC particles. Since these 
particles can efficiently decouple from Jupiter by PR drag, a large fraction of the total mass is distributed 
relatively close to the Sun. [For reference, we find that 53\%, 19\% and 3.7\% of $D=10$, 100 and 1000 $\mu$m 
particles released by JFCs, respectively, can decouple from Jupiter.]  
Figure \ref{mass} shows the mass fraction of JFC particles contained in a sphere of radius $R$ around the
Sun. The distribution is steep for $R<5$ AU and shallower for $R>5$ AU reflecting the orbital properties of 
our model JFC population. About 30\% of JFC particles, or about $1.6\times10^{19}$ g in total mass (for 
$\rho=2$~g~cm$^{-3}$ and $D=200$ $\mu$m), are located within $R<4$ AU. Also, $\approx$10\%, or about 
$5.2\times10^{18}$ g, has $R<2$ AU. 

For a comparison, assuming that the asteroidal particles with $D=200$ $\mu$m and $\rho=2$~g~cm$^{-3}$ contribute 
by 15\% to the near-ecliptic MIR fluxes, we find that the total masses in asteroidal particles with $R<2$ AU 
and $R<4$ AU are $5.3\times10^{17}$ g and $1.3\times10^{18}$ g, respectively. Thus, the total mass (or number) 
ratio of JFC to asteroidal particles in the inner solar system is $\lesssim$10. Note that this estimate applies
as far as the size distributions of JFC and asteroidal particles in the zodiacal cloud are similar, which is 
expected because both particle population live in the common collisional environment and have similar PR drag
timescales.
\subsection{Mass Influx on Earth}
We used the \"Opik algorithm (\"Opik, 1951; Wetherill, 1967) to estimate the terrestrial accretion rate of JFC 
particles expected from our model. For $30<D<300$ $\mu$m, the average impact probability of JFC particles on the 
Earth is $\sim$$5\times10^{-9}$ yr$^{-1}$ per one particle in the zodiacal cloud. A similar value 
is obtained if the impact probability is estimated from the number of direct impacts recorded by the 
$N$-body integrator. Thus, in a steady state with $\sim$$2\times10^{19}$ g in the zodiacal cloud, we estimate 
that $\sim$$10^5$ tons of JFC particles should be accreted by the Earth annually. This is larger than 
the nominal Earth's accretion rate of 20,000-60,000 tons yr$^{-1}$ as determined from LDEF (Love and Brownlee, 
1993) and the antarctic micrometeorite record (Taylor et al., 1996).

This may imply that the real Earth's accretion rate is somewhat larger than the LDEF values. Alternatively,
the LDEF constraints may imply that the real mass of the zodiacal cloud is lower than the one estimated here. As we 
discussed above, the mass of the zodiacal cloud estimated here from the IRAS data is at least a factor of $\sim$2 
uncertain. It is thus plausible that $m_{\rm ZODY} \sim 10^{19}$ g (e.g., if $\rho=1$ g cm$^{-3}$), which 
would bring our results into a better agreement with LDEF. Additional uncertainty in these estimates is related to 
the effects of collisional disruption of particles and continuous size distribution.

For comparison, if we assumed that $D=200$-$\mu$m asteroidal particles are producing the full near-ecliptic 
MIR flux measured by IRAS, the estimated terrestrial accretion rate of asteroidal particles would be 
$1.5\times10^5$ tons yr$^{-1}$. According to NVBS06 and the results obtained here, however, the asteroidal
particles contribute by only $\sim$10\% of the near-ecliptic MIR flux. Thus, we find that the asteroid particle 
accretion rate should be $\sim$15,000 tons yr$^{-1}$, or only $\sim$15\% of the JFC particle accretion rate. 
The asteroidal particles should therefore represent a relatively minor fraction of IDPs and micrometeorites in 
our collections. This explains paucity of the ordinary chondritic material in the analyzed samples (see, e.g., 
Genge, 2006).

Using the same assumptions, we estimate from our model that 16,000 tons yr$^{-1}$ and 1,600 tons yr$^{-1}$ 
of JFC particles should be accreted by Mars and the Moon, respectively. The accretion rate of JFC particles
on the Moon is thus only about $\sim$2\% of the Earth's accretion rate. This corresponds to a smaller physical 
cross-section and smaller focusing factor of the Moon. The mass influx on Mars is $\sim$20\% of the Earth's 
accretion rate. For a comparison, 1,600 tons and 100 tons of asteroidal particles are expected to fall on Mars 
and the Moon annually.

Love and Brownlee (1993) found from the LDEF impact record that $D\approx200$ $\mu$m particles should carry most 
of the mass of zodiacal particles near 1 AU, while we find here that $D\approx100$ $\mu$m provides the best fit 
to IRAS observations. This slight difference may be related to some of the limitations of our model. It can also be 
real, however, because the LDEF size distribution computed by Love and Brownlee (1993) is bending from the steep slope 
at $D>300$ $\mu$m to shallow slope at $D<50$ $\mu$m. The cross-section area should therefore be dominated by smaller
particles than the mass. From Fig. 4 in Love and Brownlee (1993) we estimate that $D\approx100$ $\mu$m particles 
should indeed dominate the total cross-section area of the zodiacal cloud at 1 AU.
\subsection{Micrometeorites}
These results have implications for the origin of micrometeorites (MMs). MMs are usually classified according
to the extent of atmospheric heating they endure (e.g., Engrand and Maurette, 1998). Cosmic spherules are
fully melted objects. Scoriaceous micrometeorites are unmelted but thermally metamorphosed objects.
The fine-grained MMs and coarse-grained MMs are unmelted objects which can be distinguished on the basis
of their grain size. Based on bulk composition, carbon content, and the composition of isolated olivine
and pyroxene grains, fine-grained micrometeorites and scoriaceous MMs, which appear to be thermally 
metamorphosed fine-grained micrometeorites, are likely related to carbonaceous chondrites. It has been 
estimated that the ratio of carbonaceous to ordinary chondrite MMs is $\sim$6:1 or larger (see, e.g., Levison et al., 
2009). This stands in stark contrast to the terrestrial meteorite collection, which is dominated by ordinary 
chondrites.

A possible solution to this discrepancy is that a large fraction of the collected micrometeorites are
particles from the Jupiter-family comets. This possibility has to be seriously considered because we find here that 
the carbonaceous JFC grains should prevail, by a large factor, in the terrestrial accretion rate of micrometeoroids.
It has been suggested in the past that a possible problem with this solution is that the cometary 
particles should encounter the Earth at large velocities (e.g., Flynn 1995), so that they either burn up in the 
atmosphere or are converted into cosmic spherules. Thus, while cometary particles could produce fully melted objects
such as the cosmic spherules it was not clear whether the less thermally processed carbonaceous MMs, 
such as the fine-grained and scoriaceous MMs, may represent cometary material. 

By assuming $t_{\rm col} \gtrsim 5\times10^5$ yr for $D=100$ $\mu$m as required by IRAS observations (see \S4.2), 
we find from our model that the mean impact speed of $D\sim200$ $\mu$m JFC particles on Earth is 
$\approx$14.5~km~s$^{-1}$ (Fig. \ref{disv}; see \S4.3 for a discussion of the size dependence of impact
speed and its relevance to meteor observations). This value is only slightly higher than that of the 
asteroidal particles ($\approx$12.5~km~s$^{-1}$). 
The comparable impact speeds of JFC and asteroidal particles in our model are a consequence of PR drag which efficiently 
circularizes the orbits before they can reach 1~AU (Fig. \ref{orbel}). We thus find that the impact speeds of the JFC 
particles are low and do not pose a serious problem. Based on this result and the high terrestrial accretion rate of 
JFC particles on Earth (\S5.2), {\it we propose that the carbonaceous MMs in our collections are grains from the 
Jupiter-family comets}. A large contribution from primitive material that may have been embedded into the main 
asteroid belt according to Levison et al. (2009) is probably not needed.
\subsection{Historical Brightness}
It is believed that the main source of JFCs is the scattered trans-Neptunian disk, which should have decayed 
by a factor of $\sim$100 over the past $\sim$4 Gyr (LD97; Dones et al., 2004). If the JFC population decayed 
proportionally,  we can estimate that the ecliptic component of the zodiacal dust should have been $\sim$100 
times brighter initially that it is now. This corresponds to the near-ecliptic 25-$\mu$m flux of about 
$7\times10^3$~MJy~sr$^{-1}$.  

A different insight into the historical brightness of the zodiacal cloud can be obtained in the framework of the 
Nice model (Tsiganis et al., 2005), which is the most complete model currently available for the early 
evolution of the outer solar system. In the Nice model, the giant planets are assumed to have formed in a 
compact configuration (all were located at 5-18 AU). Slow migration was induced in these planets by gravitational 
interaction with planetesimals leaking out of a massive primordial trans-planetary disk. After a long period 
of time, most likely some 700 Myr after formation of the giant planets (Gomes et al., 2005), planets crossed a 
major mean motion resonance. This event triggered a global instability that led to a violent reorganization of the 
outer solar system. Uranus and Neptune penetrated the trans-planetary disk, scattering its inhabitants throughout the 
solar system. Finally, the interaction between the ice giants and the planetesimals damped the orbits 
of these planets, leading them to evolve onto nearly circular orbits at their current locations.

The Nice model is compelling because it can explain many of the characteristics of the outer solar system, 
(Tsiganis et al., 2005; Morbidelli et al., 2005; Nesvorn\'y et al., 2007; Levison et al., 2008; Nesvorn\'y 
and Vokrouhlick\'y, 2009). In addition, the Nice model can also provide an explanation for the Late Heavy 
Bombardment (LHB) of the Moon (Tera et al., 1974; Chapman et al., 2007) 
because the scattered inhabitants of the planetesimal disk, and main belt asteroids 
destabilized by planetary migration, would provide prodigious numbers of impactors in the inner solar system 
(Levison et al., 2001; Gomes et al., 2005).  

Assuming that the historical brightness of the zodiacal cloud was proportional to the number of primitive 
objects that were scattered into the inner solar system on JFC-like orbits, we can estimate how it changed 
over time. In the pre-LHB stage in the Nice model, the leakage rate from the planetesimal disk beyond 15 AU 
was likely not significant relative to that at LHB. We thus expect that the MIR emission from the inner zodiacal 
cloud at $R<5$ AU should have been relatively faint, except if a massive population of particles was sustained by 
collisions in the pre-LHB asteroid belt. Here we focus on the LHB and post-LHB stages.

According to Wyatt et al. (2007), the asteroidal debris disk is expected to decay by orders of magnitude from
the time of Jupiter's formation, which marked the start of the fragmentation-dominated regime in the asteroid 
belt (e.g., Bottke et al., 2005), to LHB. It thus seems unlikely that a massive population of debris could be sustained 
over 700 Myr by the collisional grinding of main belt asteroids. Instead, it has been suggested that the collisional 
grinding in the massive trans-planetary disk at $R>15$ AU should have produced strong MIR emission peaking at $\sim$100 
$\mu$m (Booth et al., 2009; hereafter B09). Being more distant the trans-planetary disk probably decayed more slowly by collisions 
than the asteroid belt. Thus, in the pre-LHB stage, the Wien side of the trans-planetary disk emission may have exceeded 
the one from the inner zodiacal cloud down to $\sim$20 $\mu$m (B09).

During the LHB, as defined by the Nice model, large numbers of outer disk planetesimals were scattered into the inner 
solar system and the inner zodiacal cloud could have become orders of magnitude brighter than it is now. To estimate how 
bright it actually was, we used simulations of the Nice model from Nesvorn\'y and Vokrouhlick\'y (2009; hereafter NV09). 
NV09 numerically tracked the orbital evolution of 4 outer planets and 27029 objects representing the outer planetesimal 
disk. The mass of the disk was set to be 35 Earth masses. In total, NV09 performed 90 different numerical integrations 
of the Nice model, only some of which ended with the correct orbits of the outer planets.

We used these successful 
simulations to determine the number of scattered objects with JFC-like orbits as a fraction of the total initial 
number of planetesimals in the trans-planetary disk. Figure \ref{scat} shows how this fraction changed over time
in one of the NV09 successful simulations. Consistently with the estimated physical lifetime of modern JFC 
(LD97) we assumed that the physical lifetime of planetesimals after reaching $q<2.5$ AU for the first 
time was $10^4$ years. Objects past their physical lifetime did not contribute to the statistic.

Immediately after the planetary instability occurred in the Nice model, the estimated fraction of planetesimals 
having JFC-like orbits was $\approx$$7\times10^{-5}$ (Fig. \ref{scat}). It then decayed to $\approx$$10^{-6}$ 
at 50 Myr after the start of LHB. Even though the NV09 simulations gradually loose resolution at later times due to the 
insufficient number of tracked particles, we can still estimate that the fraction was $\sim$$10^{-8}$ at 
500~Myr, or about 3.4 Gyr ago in absolute chronology.

Charnoz et al. (2009) and Morbidelli et al. (2009) argued, using the crater record on Iapetus and the current 
size distribution of Jupiter's Trojans, that the total number of $D>2$~km planetesimals in the pre-LHB trans-planetary 
disk was $\sim$$10^{10}$-$10^{12}$. Using this value and Fig. \ref{scat}, we find that there were 
$\sim$$7\times10^{6}$ JFCs with $D>2$ km at time of the LHB peak, $t_{\rm LHB}$, and $\sim$$10^{5}$ JFCs at 
$t_{\rm LHB}+50$ Myr. These estimates are at least an order of magnitude uncertain mainly due to the poorly known 
size distribution of small planetesimals in the trans-planetary disk.

For comparison, Di Sisto et al. (2009) found, in a good agreement with the previous estimates of LD97, that 
there are $\approx$100 JFCs with $D>2$ km and $q<2.5$ AU in the current solar system (with about a factor of 50\% 
uncertainty in this value). Therefore, if the inner zodiacal cloud brightness reflects variations in the size of 
the historical JFC population, we find that it has been $\sim$$7\times10^{4}$ brighter at $t_{\rm LHB}$ and 
$2\times10^{3}$ brighter at $t_{\rm LHB}+50$ Myr than it is now. This would correspond to the near-ecliptic 25-$\mu$m 
fluxes of $5\times10^6$ and $10^5$~MJy~sr$^{-1}$, respectively. These values largely exceed those expected from
dust particles that were scattered from the trans-planetary disk (B09). Most of the action was apparently 
over by $t_{\rm LHB}+500$ Myr, when our model suggests that the inner zodiacal cloud was only $\sim$10 times 
brighter than it is 
now.\footnote{These estimates should only be taken as a rough guideline to the historical zodiacal cloud brightness 
because the collisional environment in the dense disk of JFC particles at LHB must have been very different from the 
one existing today. It is therefore not exactly correct to assume that the historical brightness of the zodiacal cloud 
was strictly proportional to the population of JFCs.}
\subsection{Distant Observations of the Zodiacal Cloud}
Figure \ref{zod} shows how the present zodiacal cloud would look like for a distant observer. If seen from the 
side, the brightest inner part of the zodiacal cloud has a disk-like shape with a $\approx$1.6 ratio
between the ecliptic and polar dimensions. Similar shapes have been reported by Hahn et al. (2002) from 
Clementine observations of scattered light. At a larger distance from the Sun, the shape of the 
zodiacal cloud resembles that of a walnut. The axial ratio becomes $\approx$1.3 at $R=5$ AU. 

The radial brightness profiles in Fig. \ref{zod} show a steep dimming of the zodiacal cloud with $R$. 
For $R<1$ AU, a factor of $\sim$10 in brightness is lost per 1 AU. For $1<R<5$ AU, factor $\sim$10  
is lost per 2 AU. These profiles are approximate because we ignored the effect of collisions in our model,
which should be especially important for $R\lesssim1$ AU. It is unclear how the shape of the zodiacal 
cloud would look for $R>5$ AU because we did not model the contribution from KB dust.


Figure \ref{sed} shows the Spectral Energy Distribution (SED) for distant unresolved observations of the 
zodiacal cloud. At a distance of 10 pc from the Sun, SED of the present inner zodiacal cloud is 
$1.4\times10^{-4}$ Jy at 24 $\mu$m and $5.5\times10^{-5}$ Jy at 70 $\mu$m, corresponding to the 
excesses over the Sun's photospheric emission at these wavelengths of about $3.4\times10^{-4}$ and 
$1.1\times10^{-3}$, respectively. For comparison, the approximate 3$\sigma$ excess detection limits of 
Spitzer telescope observations of Sun-like stars are 0.054 at 24~$\mu$m and 0.55 at 70~$\mu$m 
(Carpenter et al., 2009). The MIR emission of the present inner zodiacal cloud is 
therefore undetectable by distant unresolved observations with a Spitzer-class telescope. 
Specifically, the detectable emission levels are $\approx$160 and $\approx$500 larger at 24 and 70 $\mu$m, 
respectively, than those of the present inner zodiacal cloud. 

When the flux is integrated over wavelengths, we find that the fractional bolometric luminosity of the 
inner zodiacal cloud, $L_{\rm ZODY}$, relative to that of the Sun, $L_{\odot}=3.839\times10^{26}$~W, 
is $L_{\rm ZODY}/L_{\odot}\sim 2\times10^{-7}$. This is a larger value than $10^{-8}$-$10^{-7}$ 
suggested by Dermott et al. (2002) and perhaps comparable to that of KB dust at $>30$ AU (Stern et al. 1996).
The effective blackbody temperature of the zodiacal cloud can be estimated from $T_{\rm eff}=
5100/\lambda_{\rm max}$, where $\lambda_{\rm max}$ is the wavelength of the SED maximum in microns. 
With $\lambda_{\rm max}=18.5$ $\mu$m, this gives this gives $T_{\rm eff}=276$ K.
\subsection{MIR Excess during LHB}
B09 studied how the MIR excess of the solar system debris disk varied with time. According 
to them, the main source of the pre-LHB MIR emission should have been the population of dust particles 
produced by collisions in the massive trans-planetary disk at $R>20$ AU. In Fig. \ref{sed}, we show the 
model SED produced by the B09 trans-planetary disk. Being dominated by collisions (as opposed to PR-drag 
regime;  see Wyatt, 2005), the trans-planetary particles are destroyed before they could evolve 
to $R < 20$~AU. The SED emission therefore peaks at longer wavelengths ($\approx$100 $\mu$m) than the SED of 
the present zodiacal cloud ($\approx$20 $\mu$m). Also, with $\sim$35 Earth masses in the pre-LHB trans-planetary 
disk, its estimated MIR emission is strong and produces excesses of $\sim$0.1 at 24 $\mu$m and $\sim$50 at 70 
$\mu$m over the Sun's photospheric emission at these wavelengths. These values are comparable to those
of observed exozodiacal debris disks (B09).

The trans-planetary disk objects, including small dust particles, became scattered all around the solar 
system during LHB. This led to a significant depletion of the trans-planetary particle population which 
could not have been compensated by the collisional cascade because collisions became increasingly rare in the 
depleted disk.
The MIR excess should have thus dropped by orders of magnitude within several hundred Myr after the LHB start. 
B09 estimated that the 24-$\mu$m excess of dust particles scattered from the 
trans-planetary disk should have dropped to $\sim$$3\times10^{-5}$ at the present time. This is about an order 
of magnitude lower value than the 24 $\mu$m excess estimated by us for the current inner zodiacal cloud. Thus, 
there must have been a transition epoch some time after LHB when the 24 $\mu$m excess stopped being dominated 
by dust particles scattered from the trans-planetary disk and became dominated by particles produced by JFCs.

Figure \ref{lhb} illustrates how the MIR emission of the inner zodiacal cloud should have varied with time during 
LHB. The size of the JFC population was estimated by using the methods described in \S5.4. We then scaled up the 
MIR emission of the present inner zodiacal cloud by the appropriate factor (see \S5.4). We found that the 
24 $\mu$m excess reached $F_{\rm ZODY}/F_{\rm Sun}(24\mu{\rm m}) \sim 20$ at the LHB peak and stayed for 
about 100 Myr above the Spitzer's 3$\sigma$ detection limit. It dropped down to 
$\sim$10 times the value of the present zodiacal cloud at $t_{\rm LHB}+500$ Myr. We were unable to 
determine how this trend continues after $t_{\rm LHB}+500$~Myr because of the resolution issues with the NV09
simulations (\S5.4).  We expect that $F_{\rm ZODY}/F_{\rm Sun}(24\mu{\rm m})$ should have decayed by an 
additional factor of $\sim$10 from $t_0+500$ Myr to the present time. 

The 70-$\mu$m excess behaves similarly (Fig. \ref{lhb}b). It reaches $F_{\rm ZODY}/F_{\rm Sun}(70\mu{\rm m}) 
\sim 70$ at the LHB peak and decays at later times. Given the tighter detection limit of Spitzer at 70~$\mu$m, 
the 70-$\mu$m excess would remain detectable by Spitzer for 50 Myr, which is roughly half of the interval 
during which the 24-$\mu$m excess could be detected. Also, when these values are compared to the ones
estimated by B09 for the pre-LHB trans-planetary disk, we find that the 70-$\mu$m excess 
expected for JFC particles at the LHB peak is comparable to that of the pre-LHB excess produced by collisions
in the trans-planetary disk. We would thus expect that the solar system's LHB has not produced 
any significant increase of the 70-$\mu$m emission. 

Conversely, the 24-$\mu$m excess raises a factor of $\sim$100 above the B09 pre-LHB level, indicating that 
the solar system became significantly brighter during LHB at these shorter wavelengths. During LHB, the emission from 
JFC particles should have exceeded that of the scattered trans-planetary particles by $\sim$20 (compare Fig. \ref{lhb}a 
with Fig. 5 in B09). Thus, the system does not need to have a significant cold dust disk at the same time 
as the hot dust disk to provide material for the hot disk. In fact, we find here that the hot disk can be fed 
by $D\gtrsim1$ km objects, which have little total cross-section area to be detected in the cold disk, but large 
mass to sustain the hot disk upon their disintegration at $<$10 AU (see \S6). Since the decay rates of both 
populations after LHB should have been similar, their ratio should have remained roughly constant over time suggesting 
that the trans-planetary dust did not represent any significant contribution to the post-LHB emission of the 
zodiacal cloud at 24 $\mu$m.
\subsection{Debris Disks}
These results have interesting implications for our understanding of hot debris disks observed within 10 AU 
around mature stars (Trilling et al., 2008; Meyer et al., 2008; Carpenter et al., 2009; see also a review by
Wyatt 2008). It has been argued that 
some the observed brightest hot disks, such as HD 69830, HD 72905, HD 23514, $\eta$ Corvi and BD+20307, cannot be 
explained by assuming that  they are produced by the collisional grinding of the local population of asteroids 
(Wyatt et al., 2007). Specifically, Wyatt et al. (2007) pointed out that the emission from a locally produced 
population of debris is expected to be much weaker than the observed emission because disks become depleted over time 
by collisions. This problem cannot be resolved by assuming a more massive initial population because the massive population
would  decay faster. Instead, Wyatt et al. proposed that the bright hot debris disks can be seen around stars with 
planetary systems that are undergoing the LHB instability akin to that invoked in the Nice model. 

In \S5.6, we estimated how the 24-$\mu$m and 70-$\mu$m excesses varied during the solar system's LHB. We found that
the 24-$\mu$m excess should have rapidly risen by a large factor from the pre-LHB value and then gradually 
decayed. It would remain detectable by a Spitzer-class telescope for about 100 Myr after the LHB start, or 
$\sim$2\% of the Sun's current age. The 24-$\mu$m excess reached values $\gtrsim$10 at LHB, which is comparable to 
those of the brightest known hot disks (Wyatt et al., 2007). Conversely, the solar system's LHB has not produced a
significant increase of the 70-$\mu$m excess relative to the pre-LHB level. The 70-$\mu$m excess decayed after LHB 
and became undetectable by a Spitzer-class telescope after $\sim$50 Myr. Thus, if the timing is right, debris disks 
may show the 24-$\mu$m excess but not the 70-$\mu$m excess. This could explain systems such as HD 69830, which shows a 
large excess emission at 8-35 $\mu$m (Beichmann et al. 2005), but lacks 70-$\mu$m emission, and HD 101259 (Trilling 
et al., 2008). 


In a broader context, our study of the zodiacal cloud implies that: (1) the populations of small debris particles 
can be generated by processes that do not involve disruptive collisions (see \S6); and (2) observed hot dust 
around mature stars may not be produced from a population of objects that is native to $<$10 AU. 
Instead, in the solar system, most particles located within the orbit of Jupiter are fragments of 
planetesimals that formed at $>$15 AU. These icy objects are transported to $<$5 AU by gravitational encounters with 
the outer planets and disintegrate into small particles by disruptive splitting events (thought to occur due to 
processes such as the pressure build-up from heated volatiles or nucleus spin-up; see \S6). If these processes are 
common around stars harboring planets, the collisional paradigm in which debris disks are explained by collisions 
may be not as universal as thought before (see, e.g., Wyatt 2008). 
\subsection{LHB Accretion Rates}
If our basic assumptions are correct, large quantities of dust should have been accreted by the Moon, Earth and other 
terrestrial planets during LHB. For example, assuming $10^{14}$ g yr$^{-1}$ mean accretion rate over 100 Myr we estimate 
that $\sim$$10^{22}$ g of extraterrestrial material should have fallen on the Earth at the time of LHB (with $\sim$50\% of 
this mass accumulating in the first 10 Myr). This is $\sim$50 times more mass than the quantity accumulated 
by the Earth at its current accretion rate over 4 Gyr. The Moon should have accreted about 2\% of the Earth's value 
during LHB, or $\sim2$$\times10^{20}$~g in total over 100 Myr. These estimates are at least one order of magnitude 
uncertain.     

For comparison, the mass of large impactors estimated from the number and size distribution of lunar basins is 
$6\times10^{21}$ g (Hartmann et al., 2000). Thus, the total mass of dust deposited on the Moon during LHB should have 
been only $\sim$$1/30$ of that of the large impactors. 

The LHB is of fundamental interest in studies of the origin of life because it immediately precedes the oldest 
evidence for a biosphere (Awramik et al., 1983; Schidlowski, 1988; Mojzsis et al., 1996). The significance of our 
results in this context is that JFC dust grains can bring in unaltered primitive material from the outer solar system. 
They could potentially be the source of the earliest organic material that gave rise to life on Earth (e.g., 
Jenniskens, 2001; Jenniskens et al., 2004). [Asteroids are an important source of IDPs but they can accrete material 
from as far as $\sim$4 AU. It is not likely that organic material at such distances can survive the T Tauri wind of 
the young Sun.] 
\section{Comet Disruptions}
Today, $\approx$3.8 Gyr after LHB, the steady flux of JFCs from the outer solar system is keeping the zodiacal 
cloud at roughly constant brightness. We find from our numerical simulations that the mean dynamical lifetime 
of $D=200$ $\mu$m JFC particles is $10^6$ yr. Thus, to keep the zodiacal cloud at constant brightness, 
a continuous input of $\sim$$3.4\times10^{19}/10^6 = 3.4\times10^{13}$ g yr$^{-1}$, or roughly 1,100 kg s$^{-1}$ 
is required in our model. This estimate is robust because it is insensitive to the assumed $\rho$ and $D$ values 
of particles (i.e., lighter particles have shorter dynamical lifetimes). It neglects, however, the loss of particles 
due to the disruptive collisions. The real 
input rate should therefore be slightly larger, probably somewhere in the 1,000-1,500 kg s$^{-1}$ range. This 
is only slightly larger than 600-1000 kg s$^{-1}$ suggested by Leinert et al. (1983) from modeling of the Helios 1 
and 2 data.

For comparison, Reach et al. (2007) suggested from the Spitzer survey of cometary debris trails that the total 
meteoroid input from active short-period comets is $\sim$300 kg s$^{-1}$. This is $\sim$3-5 times lower value 
than what would be required, according to our estimate, to keep the zodiacal cloud brightness at constant 
brightness. While some of the uncertainties in our model and the Reach et al. results may be blamed for this discrepancy, 
we believe that this comparison may indicate that the trails of active comets represent only a fraction of the 
real mass loss in comets. In fact, it has been suggested that the main mass-loss mechanism in comets is  
their spontaneous (i.e., non-tidal) disruptions followed up by the progressive splitting of comet components into 
smaller fragments (e.g., Weissman, 1980; also see Chen and Jewitt, 1994; Boehnhardt, 2004; Fern\'andez, 2005; 
Jenniskens, 2006). 

The best documented case of comet fragmentation is that of sungrazers. These are small comet fragments that 
are detected because they pass very close to the Sun and are seen in backscattered light by solar telescopes 
(Sekanina and Chodas, 2004, 2005). Specific cases of JFCs that were observed to spontaneously 
split or break up into two or more components include 51P/Harrington, 73P/Schwassmann-Wachmann 3 and 141P/Machholz 
2 (Fern\'andez, 2005). Observations of these events show that there does not seem to be a correlation between the 
splitting event and orbital phase of the parent object, which provides motivation for how particles were 
released from JFCs in our model (\S3.1). 

Several fragmentation mechanisms may to explain the splitting of cometary nuclei: (1) rotational splitting when 
the centrifugal force exceeds nucleus' self-gravity and material strength; (2) splitting by thermal stress produced 
by the variable distance to the Sun; and (3) splitting by internal gas pressure caused by sublimation of subsurface 
pockets of volatile ices (e.g., CO). It has not been possible find the main culprit so far. Plausibly, several 
different mechanisms contribute and more observational constraints will be needed to distinguish between them. 
See Weissman (1980) and Boehnhardt (2004) for a discussion.

Fern\'andez (2005) compiled a list of 12 observed split JFCs. He found that the chance of JFC undergoing 
an observed splitting event is $\approx$1\% per orbital period. This should be taken as a lower limit on the actual 
number of splitting events because many are undetected. For example, Chen and Jewitt (1994) estimated 
that a comet has a $\sim$1\% chance to split per yr. Thus, over its active lifespan of about $10^4$ yr (LD97), 
typical JFC would undergo as many as $\sim$100 splitting events. These events may lead to the situation where the 
comet nucleus becomes completely dissolved into small particles. {\it The zodiacal cloud may thus plausibly be 
sustained by disintegrating Jupiter-family comets.} 

Our order-of-magnitude estimate supports this possibility because JFCs evolving into the inner solar system represent a 
continuous input of mass that is apparently large enough to compensate for the zodiacal cloud mass loss. Moreover, 
we found no evidence in this work for $t_{\rm JFC}$ values larger than the physical lifetime of active comets 
estimated in LD97. Most JFC comets should therefore be dissolved on timescales comparable to their active lifetime. 

Using the size distribution of JFCs from Tancredi et al. (2006), we find that the total mass of JFCs with radius 
$0.1<s<10$ km and $q<2.5$ AU is $3.9\times10^{14}$ g. Assuming that this mass is injected into the zodiacal cloud
every $10^4$ yr (LD97), we find the total mass input of 12,000 kg s$^{-1}$. This is significantly larger than 
the mass input required to maintain the zodiacal cloud in a steady state (1,000-1,500 kg s$^{-1}$), possibly
suggesting a $\sim$10\% yield of the disintegration process. Note, for example, that some JFCs or their large 
fragments can be removed (e.g., impact planets or leave the solar system) before they could fully disintegrate. 
Also, icy particles released by comets sublimate at $R<5$ AU and do not contribute to the inner zodiacal cloud. 

Di Sisto et al. (2009) determined the physical lifetime of JFCs to be $\sim$3 times shorter than LD97. Using Di 
Sisto et al. estimate, we find that the JFC population should require the mass input of 35,000 kg s$^{-1}$. The yield of the 
disintegration process may thus be as low as $\sim$3\%. For comparison, Di Sisto et al. found the following fractions 
of JFCs that are completely dissolved by splitting events: 51\% for radius $s=1$ km, 13\% for $s=5$ km and 8\% 
for $s=10$ km. 

The initial size distribution of particles resulting from the splitting process is uncertain, but meteor showers 
from freshly ejected dust trails, such as Phoenicids, indicate that the distribution should be fairly flat with most mass 
in mm to cm size grains. This initial size distribution is modified by collisions as JFC particles decouple from 
Jupiter and drift to lower $R$ where collisions are more common. As discussed in \S4.2, the collisional effects explain 
why $D \approx 100$ $\mu$m provide the best fit to the IRAS data, because these intermediate-size particles have longest 
lifespans (e.g., Gr\"un et al., 1985; Dermott et al., 2001).

Additional evidence that disruptions/splitting events of JFCs may dominate the population of interplanetary particles 
in short-period orbits comes from observations and modeling of the meteor showers. Specifically, it has been established 
that most meteor streams were produced by recent ($<$few thousand years ago) comet disruptions (see Jenniskens 2008 for 
a review). For example, 1956 Phoenicids and near-Earth object 2003 WY25 are most likely fragments produced by a breakup 
of D 1819 W1 (Blanpain) (Jenniskens and Lyytinen, 2005; Watanabe et al., 2006). In addition to 2P/Encke, there are other 
known comet fragments moving in the Taurid stream (Jenniskens, 2006), also pointing to
a disruption event. Geminids, Phaeton, and 2005 UD can also be linked to the common parent body (Jenniskens, 2006; Ohtsuka, 2005; 
Jewitt and Hsieh, 2006). The type of disintegration that produced these large fragments and meteoroid streams is probably 
like that of the 1995 breakup of 73P/Schwassmann-Wachmann 3, which will cause a shower of tau-Herculidis in 2022.

The meteoroid streams that were associated with comet disruptions are much stronger than the meteoroid streams 
produced by active JFCs. Thus, the strong meteoroid streams may represent an important link between JFCs and the zodiacal 
cloud. They should become increasingly more dispersed due to effects of planetary perturbations. Eventually, the particles should 
be well mixed in orbital space, producing both the sporadic meteoroid complex and zodiacal cloud. Notably, the 
time-integrated  flux of visual meteors at Earth is dominated by about a factor of $\sim$10 by sporadics (Jones and Brown, 
1993).

Based on modeling of meteor radar observations, Wiegert et al. (2009) demonstrated that the prominent helion/anti-helion 
pair of sporadic meteors is most likely produced by JFCs. This result provides further support to the zodiacal 
cloud model proposed in this work because it shows that the JFC particles are an important part of the zodiacal dust 
complex at 1 AU. According to Wiegert et al., the north/south apex pair is probably produced by retrograde long-period 
comets, perhaps suggesting an OCC component in the zodiacal cloud. As we showed in \S4, a small contribution of isotropic 
OCC particles is also required to explain IRAS observations of the zodiacal cloud. 
\section{Comparison with Previous Work}
The origin and evolution of the zodiacal cloud has been the subject of numerous studies. For example, Liou et al. 
(1995) suggested, based on modeling in many ways similar to our own, that the observed shape of the zodiacal cloud 
can be accounted for by a combination of $\sim$1/4-1/3 of asteroid dust and $\sim$2/3-3/4 cometary dust. We found a 
much larger JFC contribution and much smaller asteroid contribution in this work. The cause of this difference is 
unknown. Possibly, it stems from some of the approximations used by Liou et al. (1995). For example, they used 
particles from comet 2P/Encke to represent the whole population of particles released by JFCs. This comet has
a special orbit ($a=2.22$ AU and $q=0.33$ AU) that is not representative for the JFC population as a whole.

Different constraints on the origin of the zodiacal cloud have been obtained from modeling the asteroid dust 
bands. For example, Dermott et al. (1994b) suggested that the particles originating in the main asteroid belt supply
$\sim$1/3 of the zodiacal cloud, while NVBS06 estimated the contribution of asteroidal particles to be $<$10\%.
Our results presented in \S4 are more in line with the NVBS06 estimate. Specifically, we found that a $\gtrsim$20\% 
asteroid contribution to the {\it near-ecliptic} MIR fluxes can be ruled out from IRAS observations. If correct, 
this limits the asteroid contribution to the {\it overall} cross-section of the zodiacal cloud to a sub-10\% level.

Hahn et al. (2002) used Clementine observations of the zodiacal cloud at optical wavelengths and arguments based 
on the inclination distribution of small bodies in the solar system to argue that at least $\approx$90\% of 
the zodiacal cloud cross section enclosed by a 1-AU-radius sphere around the Sun is of cometary origin. They 
also found that $\approx$45\% optical cross-section at 1 AU comes from JFCs and/or asteroids. Unfortunately, 
a distinction between JFC and asteroid dust could not have been made because Hahn et al. used an approximate 
model for the interplanetary dust complex. According to our model, the contribution of JFC is much larger than the 
one found by Hahn et al. (2002). Thus, while we agree with the general conclusion of Hahn et al. about the 
predominant comet dust population, our results are more specifically pointing out JFCs as the main source.
\section{Origin of Particle Populations beyond Jupiter}
Our findings are in a broad agreement with the results obtained from dust detectors onboard spacecrafts. For 
example, Altobelli et al. (2007) identified two main groups of particles in the Cassini's Cosmic Dust Analyzer 
data set (measurements in the ecliptic plane between Jupiter and Saturn). The first group of 
impactors consists of particles on bound and prograde orbits, most probably having moderately 
eccentric and moderately inclined orbits. These grains are consistent with JFCs. Impactors of the 
second group were identified as small interstellar dust particles, perhaps including a minority of 
beta meteoroids. 

Landgraf et al. (2002) reported results from the dust experiments onboard the Pioneer 10 and 11 
spacecrafts. They found that the spatial number density of $\gtrsim$10 $\mu$m particles at the ecliptic is 
only slowly declining with heliocentric distance in the 3-18 AU range. Specifically, there is no obvious gap 
beyond 4 AU, expected if asteroidal particles were the dominant source of dust in the inner solar system
(Fig. \ref{mass}). 

The nearly constant spatial density of the circumsolar dust beyond 5 AU is puzzling. To explain it, Landgraf et 
al. (2002) proposed that particle populations beyond Saturn are be dominated by dust produced in KB collisions
(see also Moro-Mart\'{\i}n and Malhotra, 2003). The observed radial density profile beyond 5 AU is produced 
in their model by combining the contributions from KB particles, whose spatial density raises with $R$, and
cometary particles, whose density declines with $R$. Indeed, the spatial density of JFC particles that we obtain
from our model rather steeply declines with $R$ at $R>5$ AU. Thus, the Kuiper belt dust may indeed be needed to 
explain Pioneer measurements. [A possible caveat of these considerations is that the impact rates measured by 
Pioneer 1 and 2 should be mainly those of $\sim$10 $\mu$m particles, while the dominant size of particles in the 
inner zodiacal cloud is $\sim$100-200 $\mu$m.]

An alternative possibility is that we do not correctly determine the distribution of JFC grains for $R>5$ AU 
in our model. This alternative is attractive for the following reasons.

If the trans-Neptunian population is in the collisional equilibrium for $D<1$ km, most mass should be 
contained in comet-size and larger bodies rather than in $D<1000$ $\mu$m grains. Since the transfer of this material 
to the Jupiter-crossing orbit is size-independent (driven mainly by the encounters to outer planets), JFCs must 
represent much more mass than their grain-sized orbital counterparts. Thus, assuming that JFCs can be efficiently 
dissolved by splitting events, the dust population they produce should be much more important than the one evolving 
from the Kuiper belt in the form of dust grains. 

Di Sisto et al. (2009) found a very high splitting 
rate of JFCs with only a shallow dependence on their perihelion distance ($\propto q^\alpha$ with $\alpha \sim -0.5$).
Thus, if JFCs can be efficiently dissolved at large $q$, the radial distribution of JFC dust should significantly 
differ from the one obtained here (see \S3.1 for our assumptions). The spatial density of JFCs is proportional 
to $R^\gamma$ with $\gamma \sim 0.5$ (LD97; Di Sisto et al., 2009). Since $|\alpha| \sim \gamma$, the number of 
splitting events, and therefore the number of generated JFC particles, should be roughly independent of $R$. It 
might thus be plausible to explain the Pioneer measurements with JFC particles alone, i.e., without a major 
contribution from KB particles. A detail investigation into these issues goes beyond the scope of 
this paper.   

%
\section{Summary}
We developed models for various source populations of asteroid and cometary dust particles. These 
models were based on our current understanding of the origin and evolution of asteroids, Jupiter-family, 
Halley-type, and Oort-Cloud comets. We launched sub-mm particles from these populations and tracked 
their orbital evolution due to radiation pressure, PR drag and planetary perturbations. The thermal MIR 
emission from the synthetic particle distributions were determined and the results were compared to IRAS 
observations. 

\begin{figure}
\epsscale{0.5}
\plotone{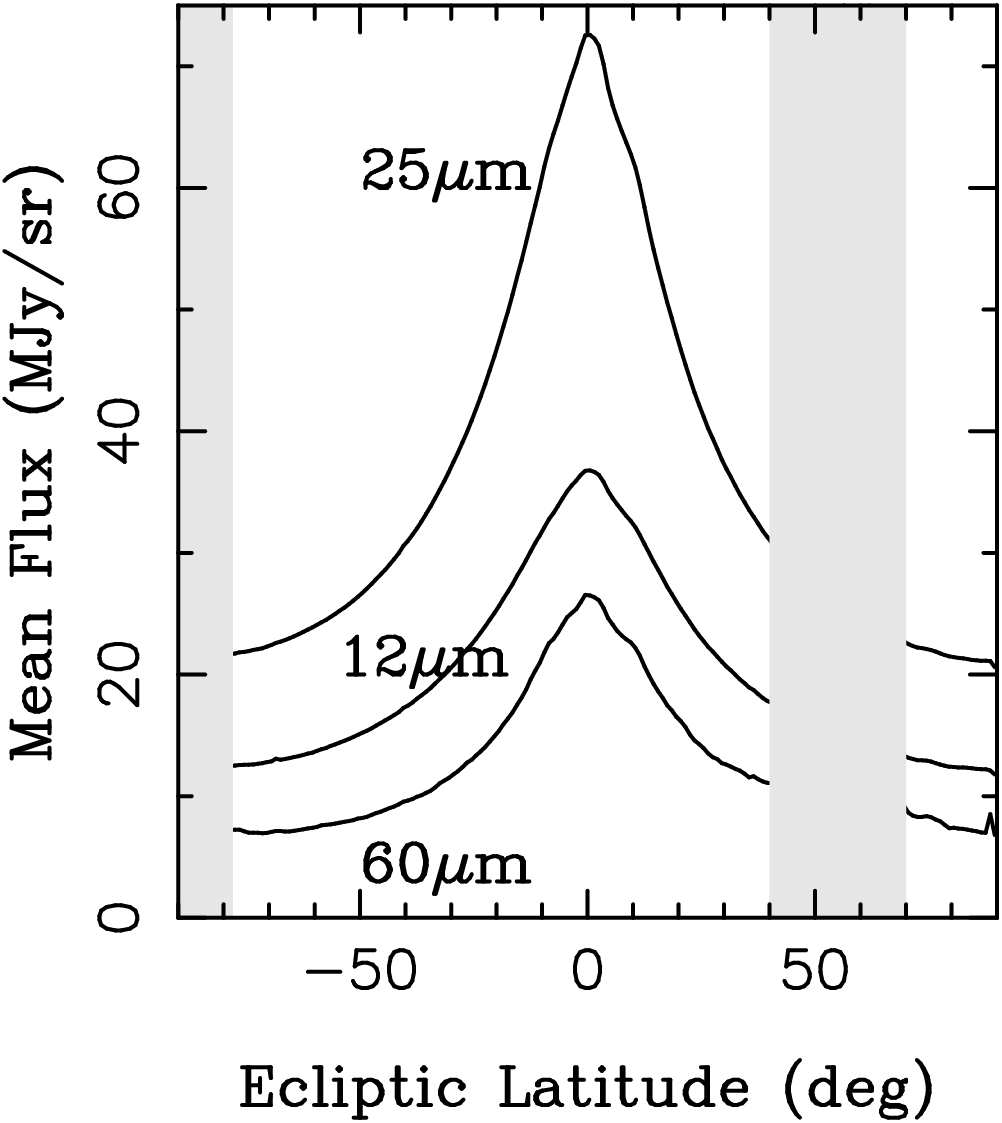}
\caption{Mean IRAS profiles at 12, 25 and 60 $\mu$m wavelengths. To make these profiles, the selected 
IRAS scans were centered at the ecliptic, smoothed by a low-pass filter, and combined together. The gray
rectangles at $l<-78^\circ$ and $40^\circ<l<70^\circ$ block the latitude range where the mean fluxes 
were significantly affected by the galactic plane emission. We do not use the excluded range in this
work. The uncertainties of the mean flux values are not shown here for clarity; they are too small 
to clearly appear in the plot. The characteristic errors at different wavelengths averaged over latitudes 
are $\sigma_{12\mu {\rm m}}=0.59$ MJy sr$^{-1}$, $\sigma_{25\mu {\rm m}}=1.1$ MJy sr$^{-1}$ and 
$\sigma_{60\mu {\rm m}}=2.7$ MJy sr$^{-1}$. 
They increase with wavelength due to the larger role of galactic emission at longer wavelengths.}
\label{mean}
\end{figure}

\begin{figure}
\epsscale{0.6}
\plotone{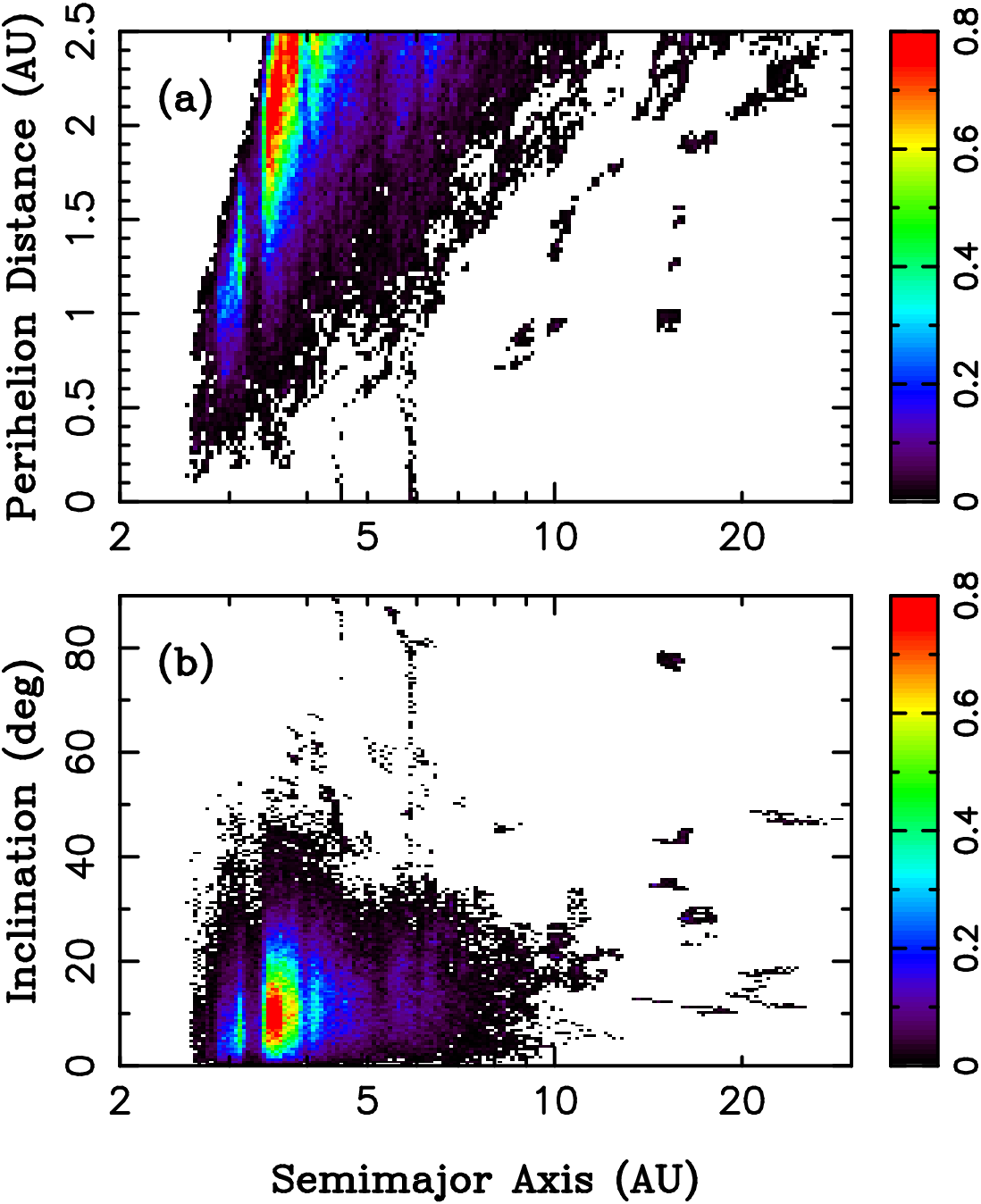}
\caption{Orbital distribution of JFCs from the LD97 model. Panels show the perihelion 
distance (a), and inclination (b), as functions of the semimajor axis for JFCs with $q<2.5$ AU 
and $t_{\rm JFC}=12,000$ yr. See \S3.1 for the definition of $t_{\rm JFC}$. The 2:1 and 3:2 mean 
motion resonances with Jupiter correspond to the gaps in the distribution at $a\approx3.3$ and 3.96 AU,
respectively. The inclination 
distribution of JFCs shown here is remarkably similar to that obtained by Di Sisto et al. (2009; 
their Fig. 10).} 
\label{bin1}
\end{figure}

\begin{figure}
\epsscale{0.5}
\plotone{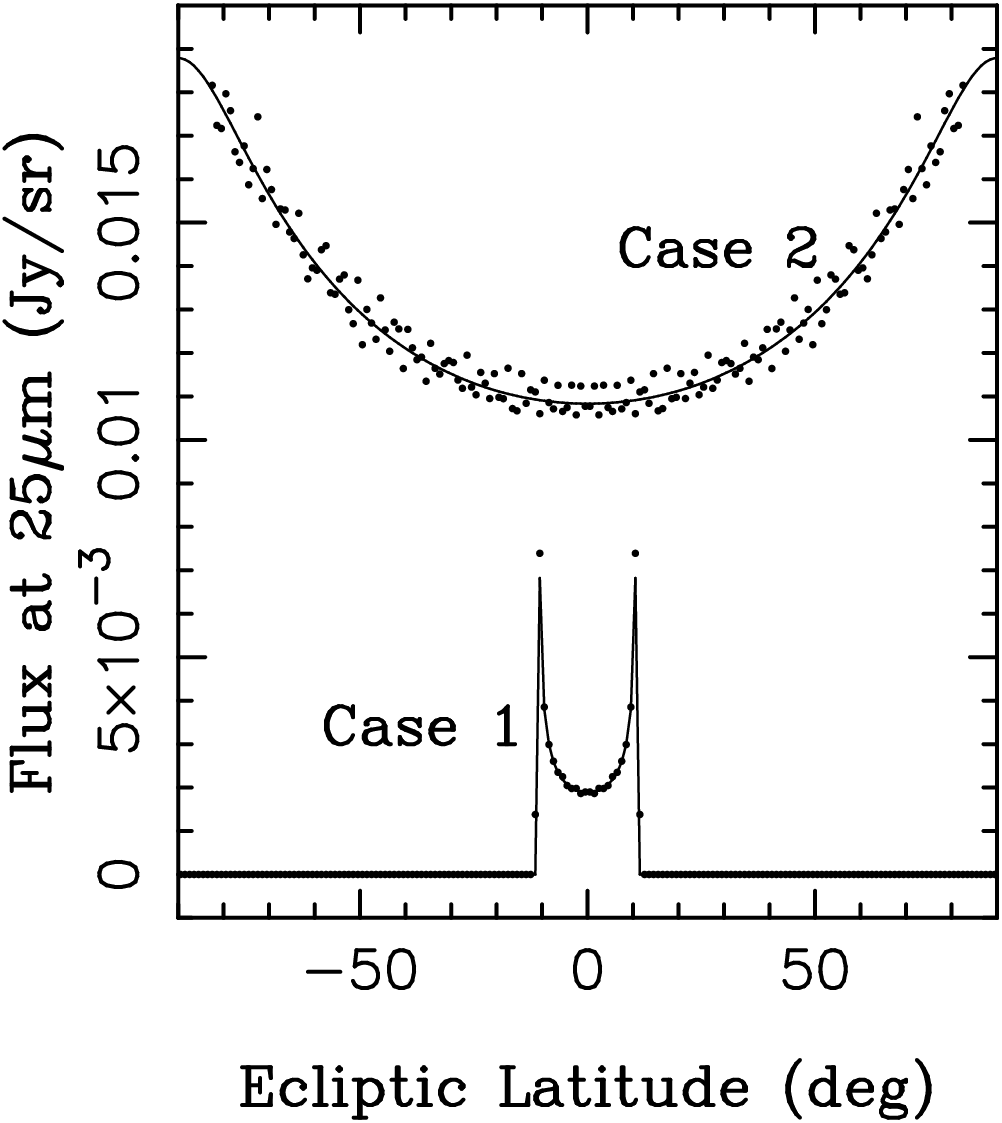}
\caption{Comparison between results obtained from the particle algorithm (dots) and SIRT (solid lines).
Case 1 corresponds to asteroidal particles with $a=2.5$ AU, $e=0.1$ and $i=10^\circ$. 
Case 2 corresponds to cometary particles crossing the Earth's orbit with $a=1$ AU, $e=0.5$ and 
$i=50^\circ$. In both cases we assumed that particles have $D=100$ $\mu$m and are distributed
randomly in $\Omega$, $\varpi$ and $M$. The flux at 25 $\mu$m was normalized to a population
of $10^{15}$ particles in Case 1 and $2\times10^{15}$ particles in Case 2. Observations with 
$r_{\rm t}=1$ AU and $l_\odot=90^\circ$ vere assumed. In Case 2, the particle algorithm shows a scatter 
around the exact solution due to the rough resolution of the distribution near the telescope's 
location. We used $5\times10^{10}$ orbit clones in the particle algorithm.}
\label{check}
\end{figure}

\begin{figure}
\epsscale{0.5}
\plotone{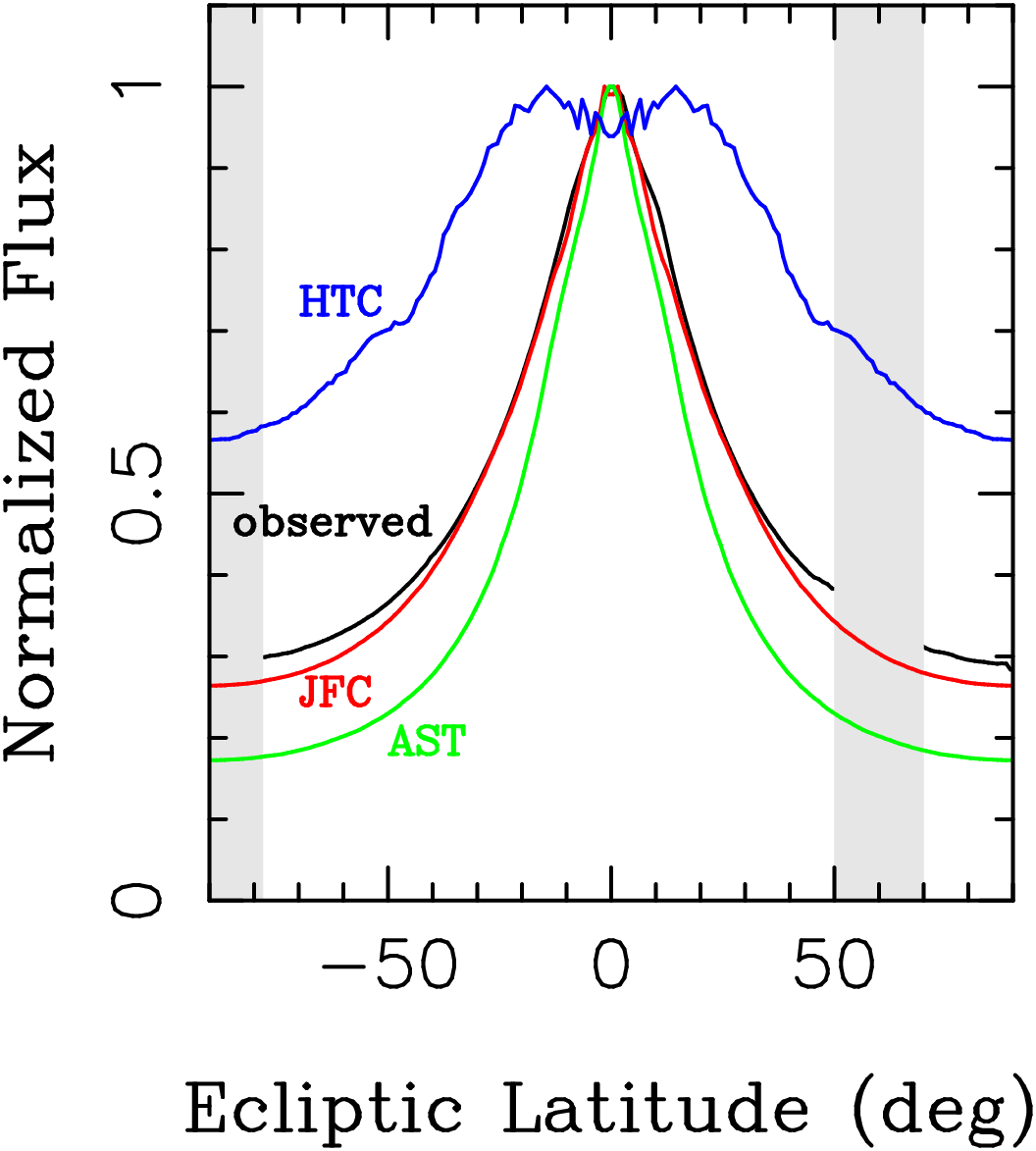}
\caption{Comparison of the 25-$\mu$m profiles produced by different sources with IRAS observations. 
The black line shows our mean IRAS scan for $l_\odot=90^\circ$. The colored lines show 
profiles expected from different source populations: asteroids (green), JFCs (red) and HTCs (blue).
The OCC flux, not shown here for clarity, is a nearly constant function of latitude.
The maximum flux in each profile has 
been normalized to 1. We used $D=100$ $\mu$m and $t_{\rm JFC}=12,000$~yr. The main differences between 
profiles are not sensitive to the exact choice of $D$ and other model parameters.}
\label{mod1}
\end{figure}

\begin{figure}
\epsscale{0.5}
\plotone{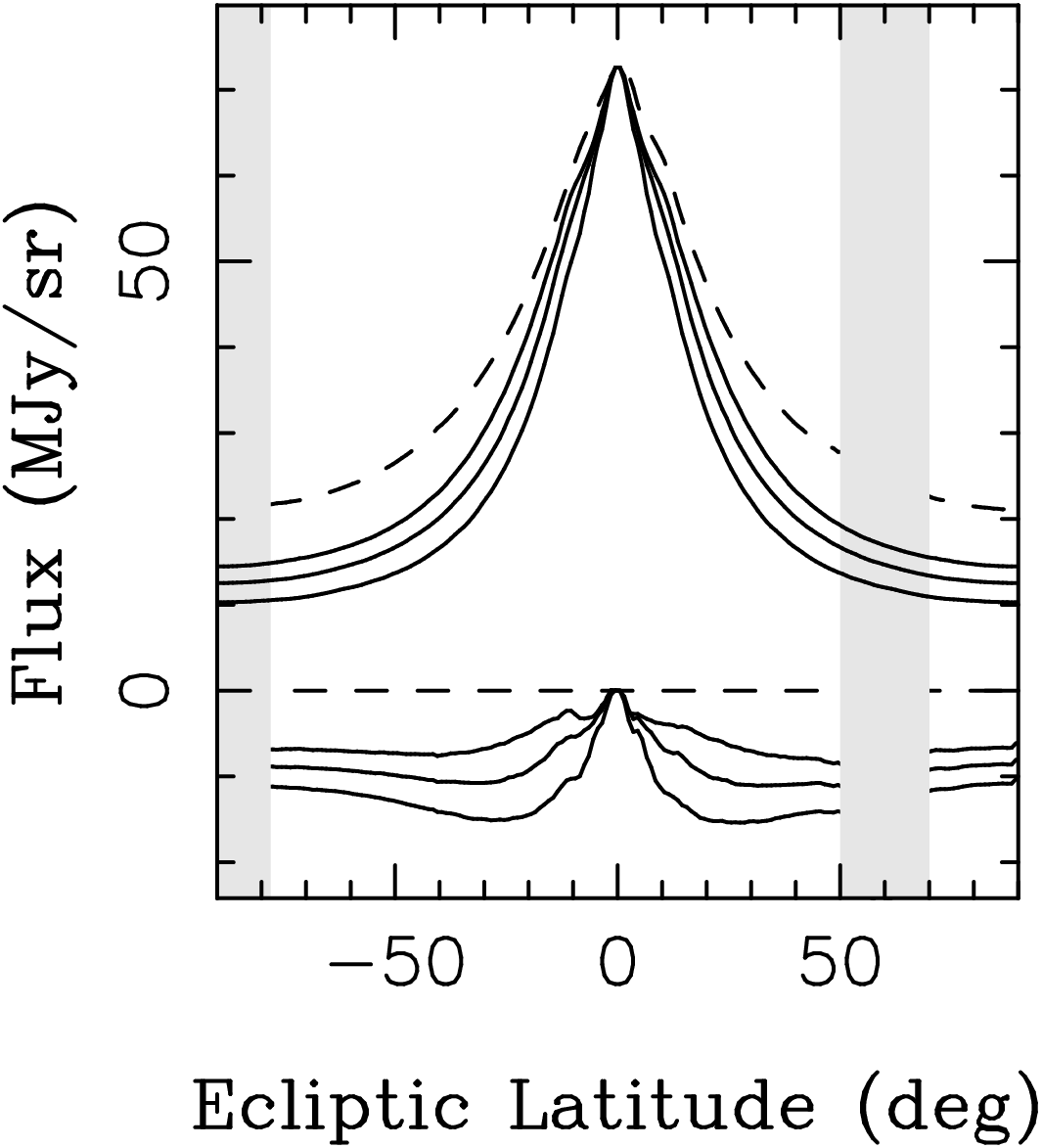}
\caption{Dependence of the shape of 25-$\mu$m profiles produced by asteroidal particles on $D$.
The dashed line shows the mean 25-$\mu$m IRAS profile for $l_\odot=90^\circ$. The upper solid curves show 
the model results for the same wavelength and elongation. The bottom lines show the residual 
flux obtained by substracting the model flux from the mean IRAS profile. Results for $D=30$, 100 and 
300 $\mu$m asteroidal particles are shown with slightly broader profiles corresponding to larger $D$.
The profiles for $D=10$ and 1000 $\mu$m, not shown here, are narrower than the ones for $D=30$ $\mu$m.
For $D=1000$ $\mu$m, this is mainly due to the effects of disruptive collisions that destroy large grains 
before they could evolve down to 1 AU (see discussion in \S4.2). None of the model profiles obtained with 
asteroidal particles can match IRAS observations.} 
\label{ast1}
\end{figure}

\begin{figure}
\epsscale{0.4}
\plotone{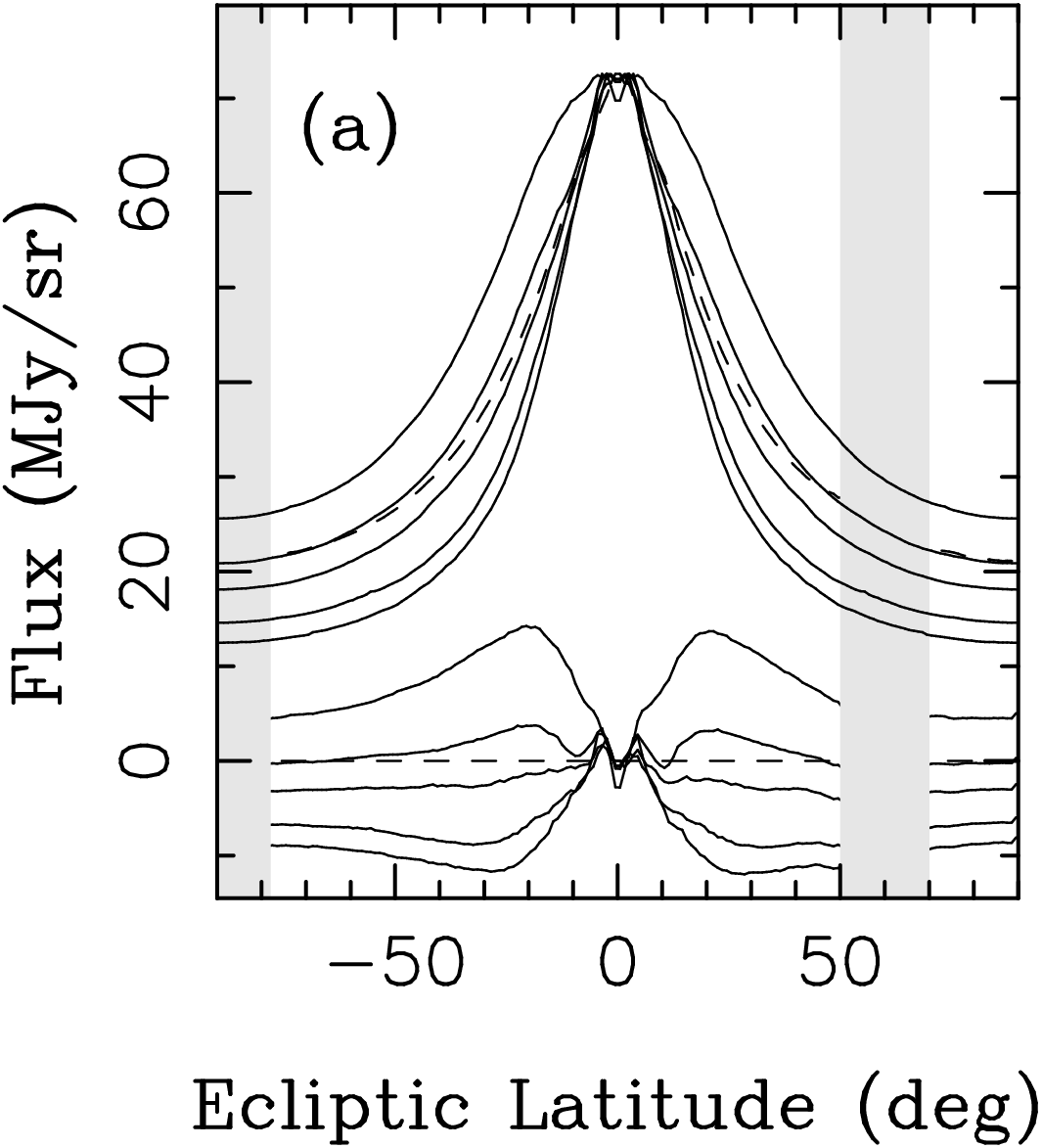}\hspace*{5.mm}
\plotone{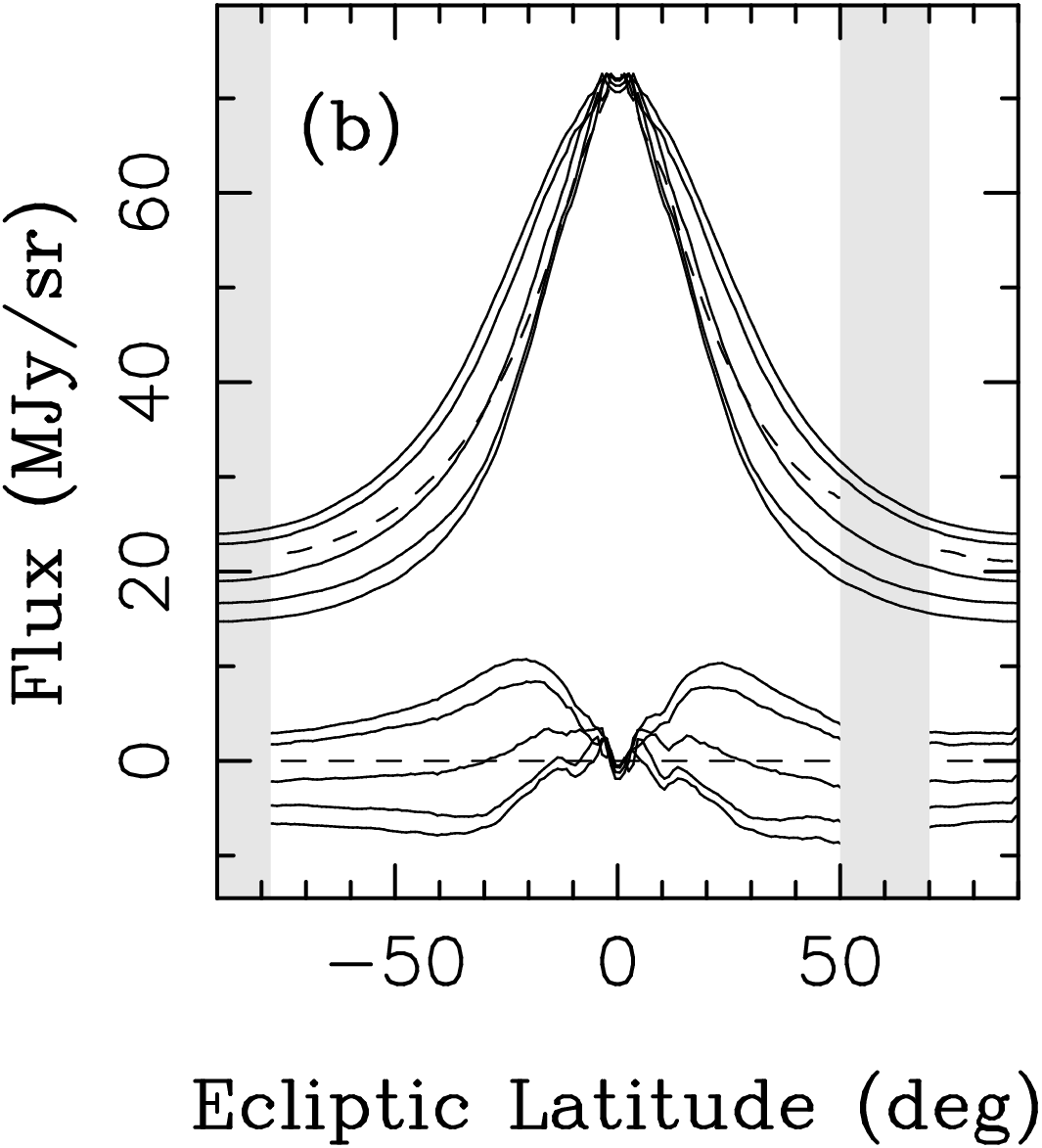}\vspace*{5.mm}
\plotone{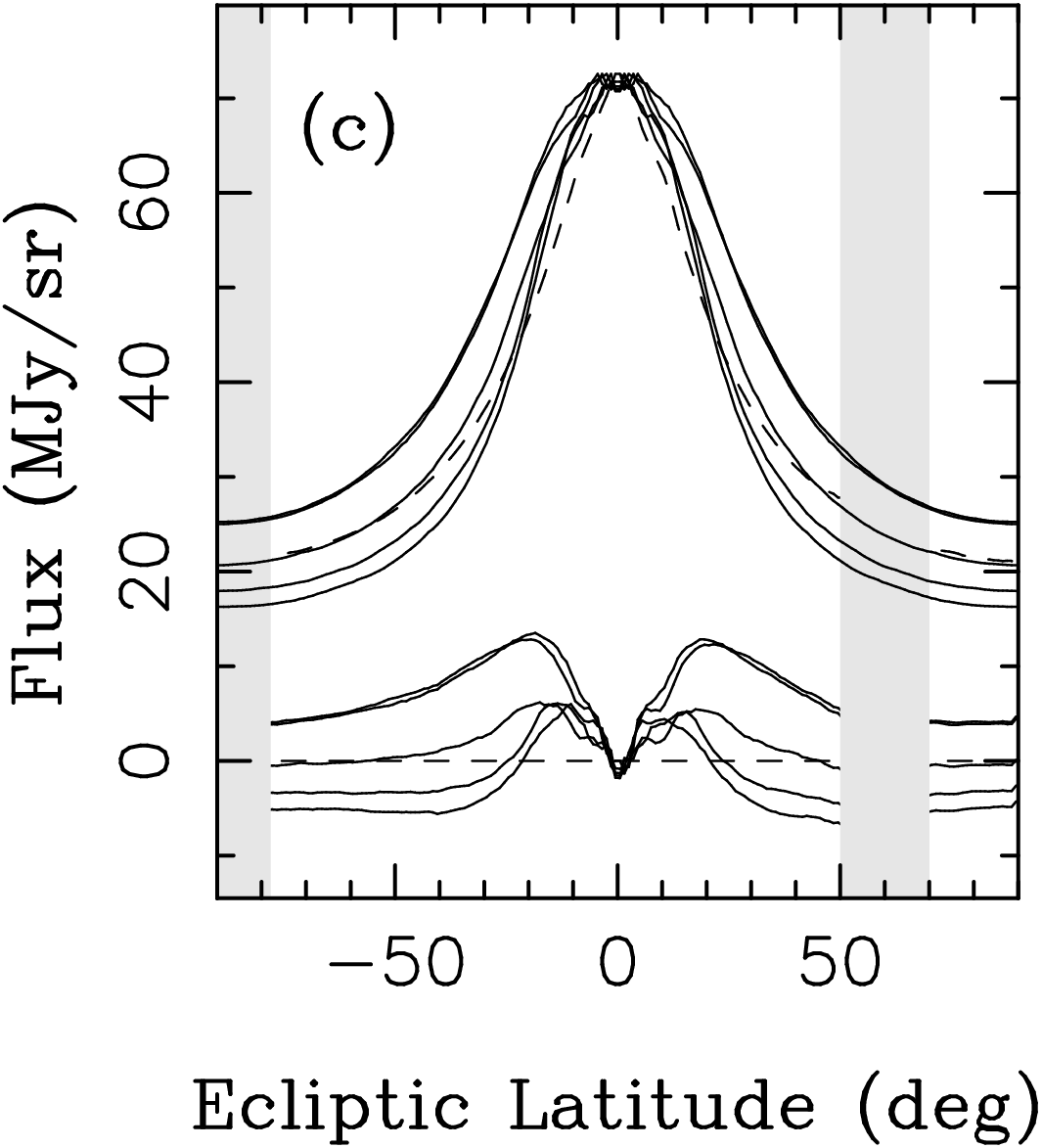}\hspace*{5.mm}
\plotone{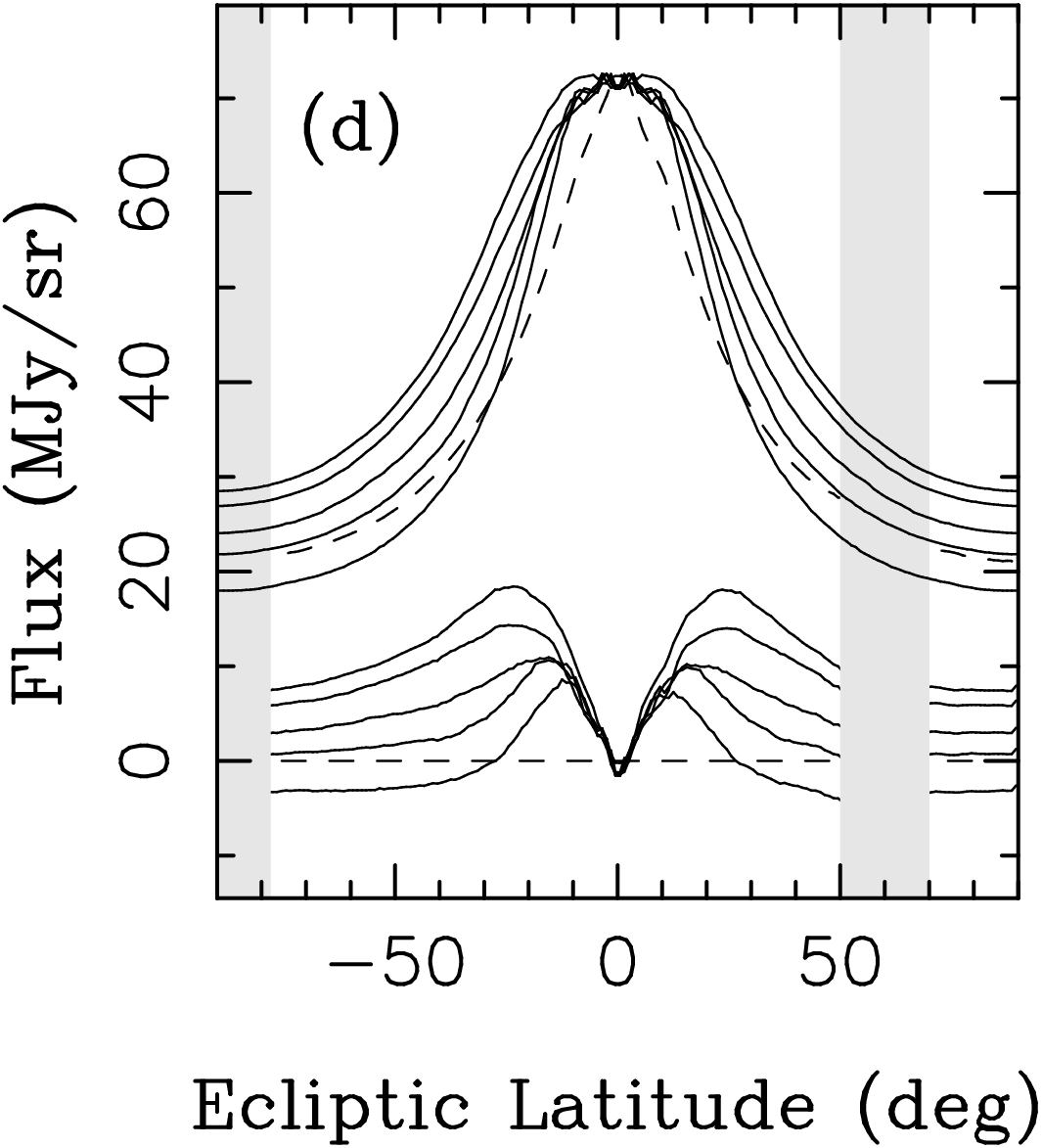}
\caption{Dependence of the shape of 25-$\mu$m profiles produced by JFC particles on $D$ and $t_{\rm JFC}$.
The panels show results for different $t_{\rm JFC}$: (a) $t_{\rm JFC}=
12,000$ yr, (b) $t_{\rm JFC}=30,000$ yr, (c) $t_{\rm JFC}=50,000$ yr and (d) $t_{\rm JFC}=100,000$ yr.
The dashed line in each panel shows the mean 25-$\mu$m IRAS profile for $l_\odot=90^\circ$. 
The upper solid curves show the model results for the same elongation. The bottom lines show the residual flux 
obtained by substracting the model fluxes from the mean IRAS profile. Results for $D=10$, 30, 100, 300 and 1000 
$\mu$m are shown in each panel with broader profiles corresponding to larger $D$.  Some of these model 
profiles do not match IRAS observations well. Specifically, $t_{\rm JFC}>50,000$ yr, $D>300$~$\mu$m 
and $D=10$ $\mu$m can be clearly ruled out.} 
\label{dep}
\end{figure}

\begin{figure}
\epsscale{0.4}
\plotone{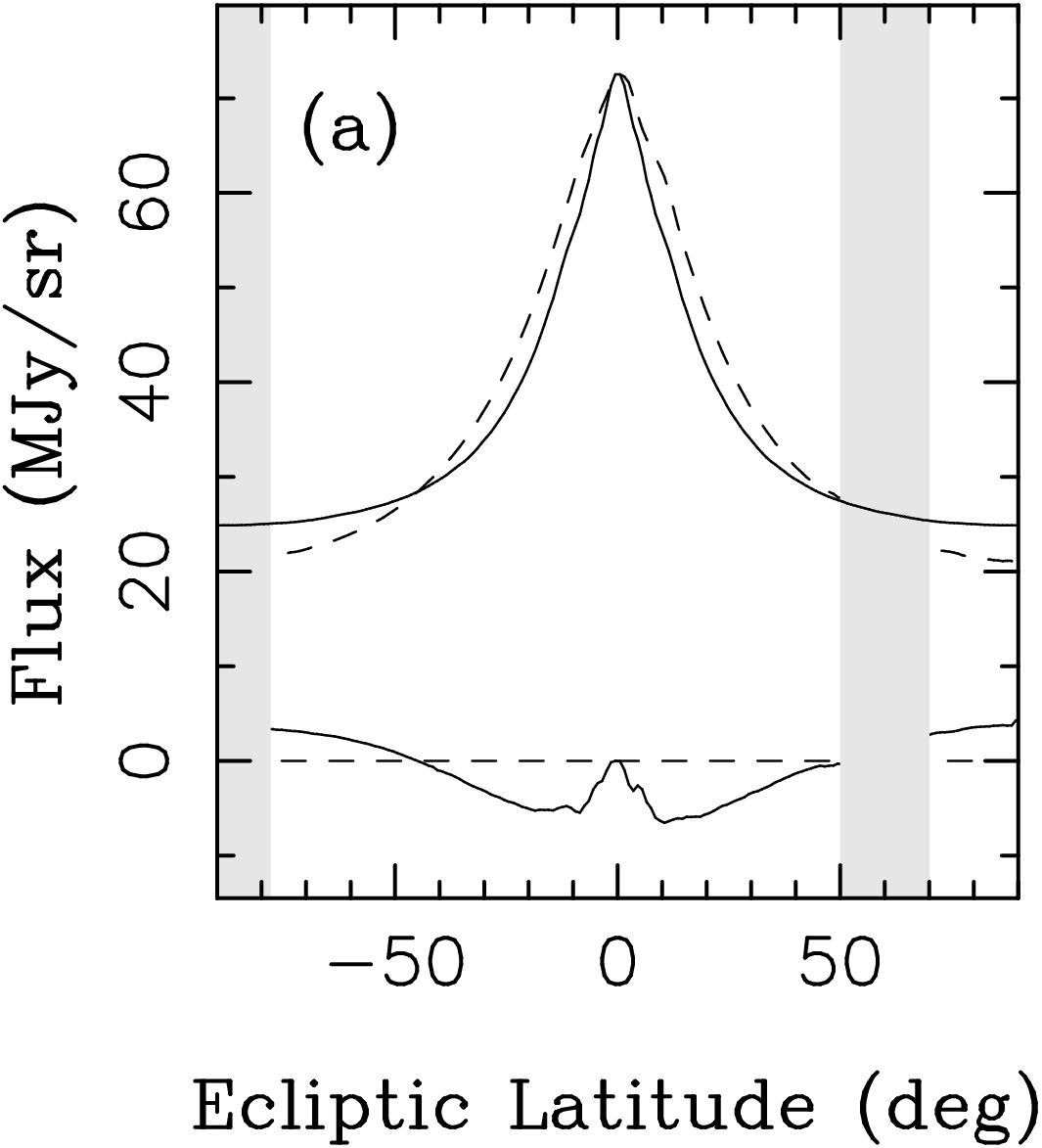}\hspace*{5.mm}
\plotone{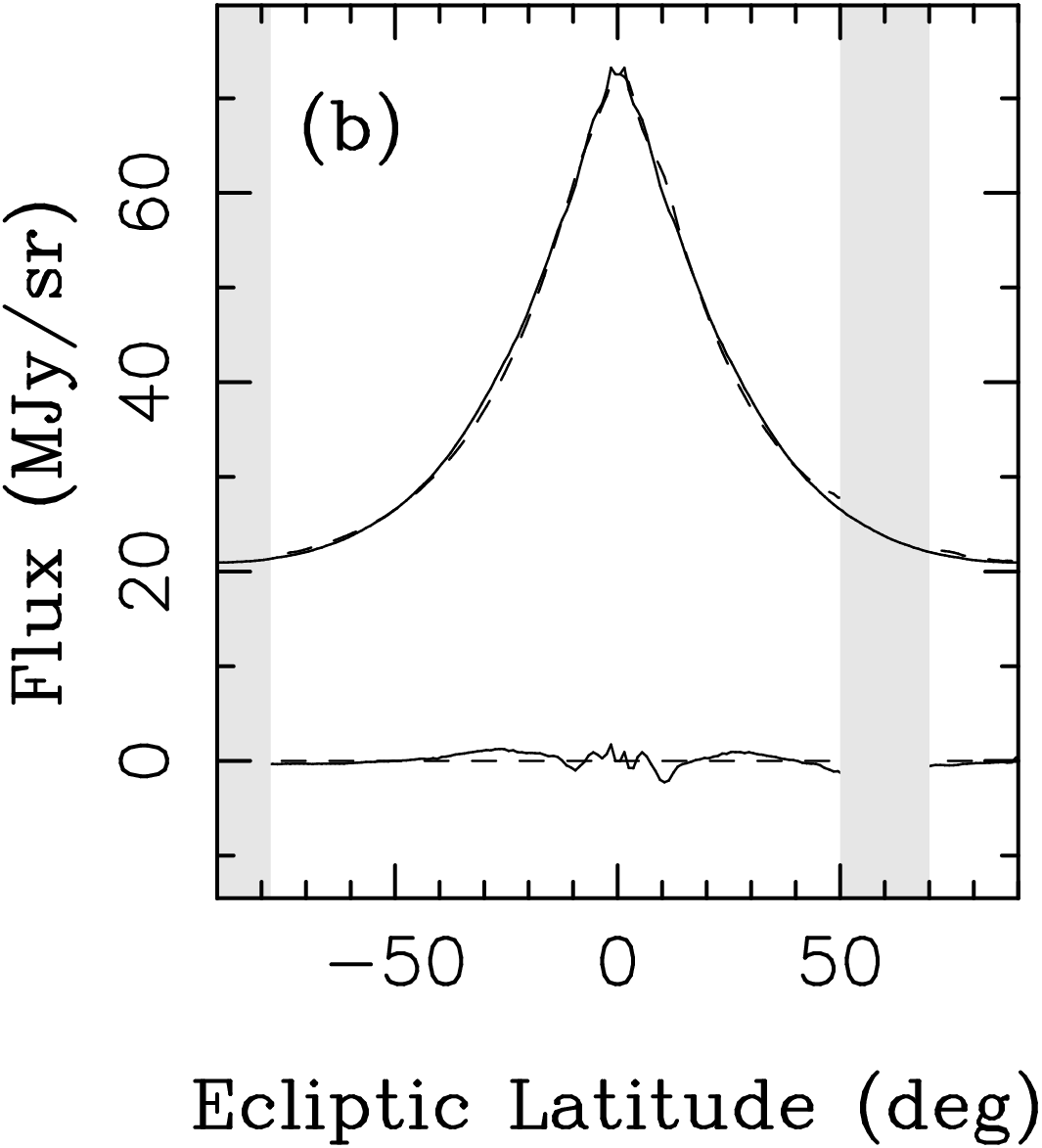}
\caption{Examples of fits where we modeled the zodiacal cloud as having two sources. Fluxes at
25 $\mu$m are shown. (a) The best-fit model with asteroid and OCC sources. This model does not fit 
IRAS observations well. The model profile is too narrow near the ecliptic and too wide overall. 
(b) Our best two-source model. Here we used $\alpha_{\rm JFC}=0.97$, $\alpha_{\rm OCC}=0.03$, $D=100$ 
$\mu$m and $t_{\rm JFC}=12,000$ yr. The faint isotropic component improves the fit quality so that 
$\eta^2=0.36$ in (b). This may suggest that the zodiacal cloud contains a small but significant fraction of 
OCCs particles.} 
\label{fit2}
\end{figure}

\begin{figure}
\epsscale{0.6}
\plotone{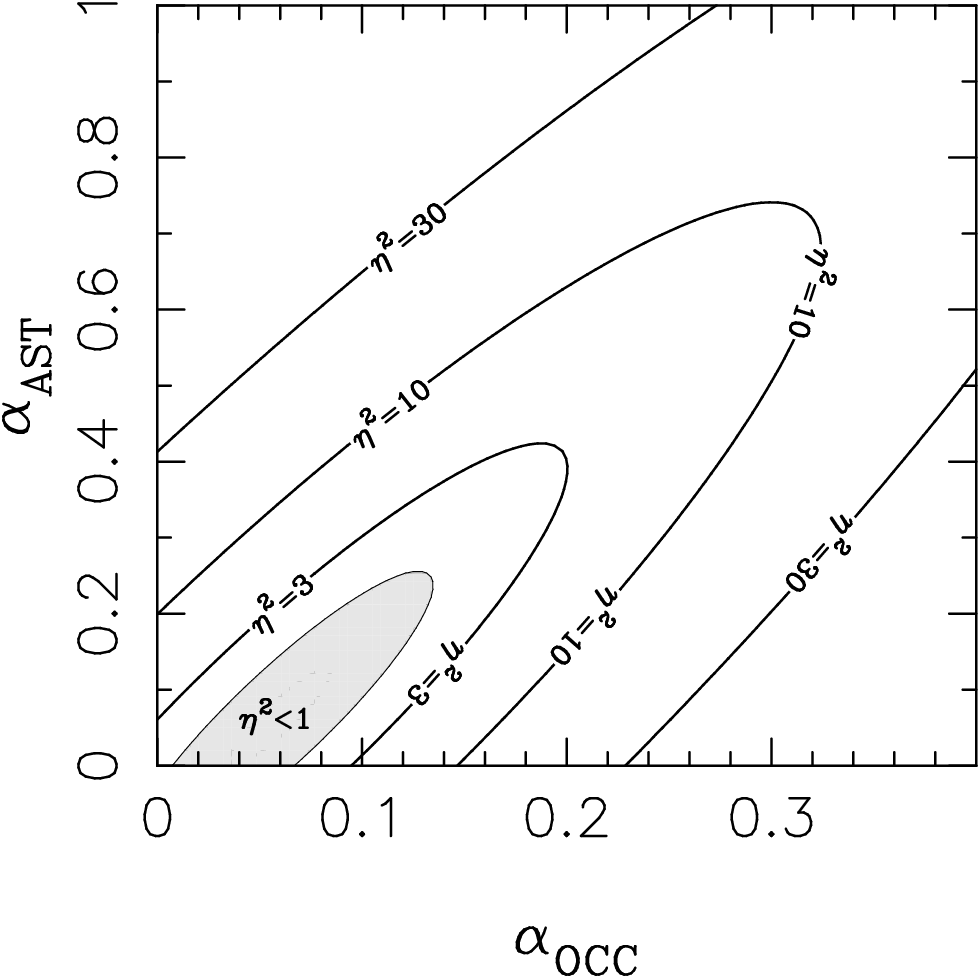}
\caption{Model constraints on the contribution of asteroid and OCC particles to the zodiacal cloud. Here we 
used a three-source model with $D=100$ $\mu$m and $\alpha_{\rm HTC}=0$. For a range of the $\alpha_{\rm OCC}$ 
and $\alpha_{\rm AST}$ values, we set $\alpha_{\rm JFC} = 1 - \alpha_{\rm OCC} - \alpha_{\rm AST}$, and 
calculated $\eta^2$ (Eq. \ref{eta}) for each model. The contours show $\eta^2 = 3$, 10 and 30. The shaded 
area denotes the parameters of our best-fit models with $\eta^2 < 1$. These models have $\alpha_{\rm OCC}<0.13$
and $\alpha_{\rm OCC} < 0.22$ thus placing an upper limit on the near-ecliptic contribution of asteroid and OCC 
particles.} 
\label{alphas}
\end{figure}

\begin{figure}
\epsscale{1.0}
\plotone{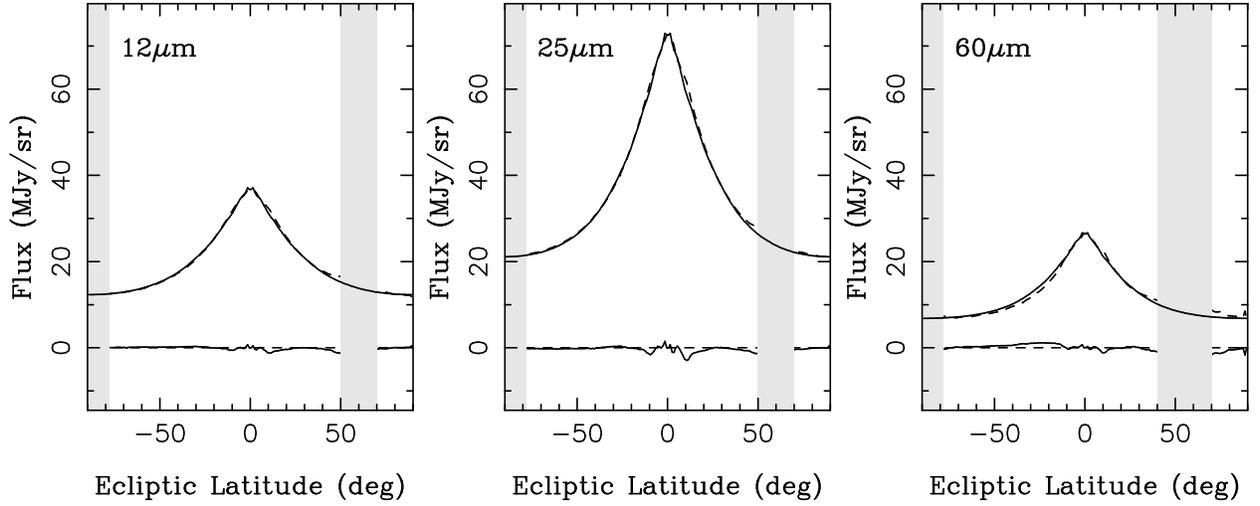}
\caption{Our preferred fit with $\alpha_{\rm JFC}=0.85$, $\alpha_{\rm OCC}=0.05$, $\alpha_{\rm AST}=0.1$ 
and $\alpha_{\rm HTC}=0$. Particles with $D=100$ $\mu$m and $t_{\rm JFC}=12,000$ yr were used here.  
The dashed lines show the mean IRAS profiles at 12, 25 and 60 $\mu$m. The upper and lower solid 
lines are the model and residual profiles, respectively. The wiggle in the residual profiles for $|b|<10^\circ$ 
may occur due to a slight problem with our asteroid dust band model.} 
\label{fit4}
\end{figure}

\begin{figure}
\epsscale{0.4}
\plotone{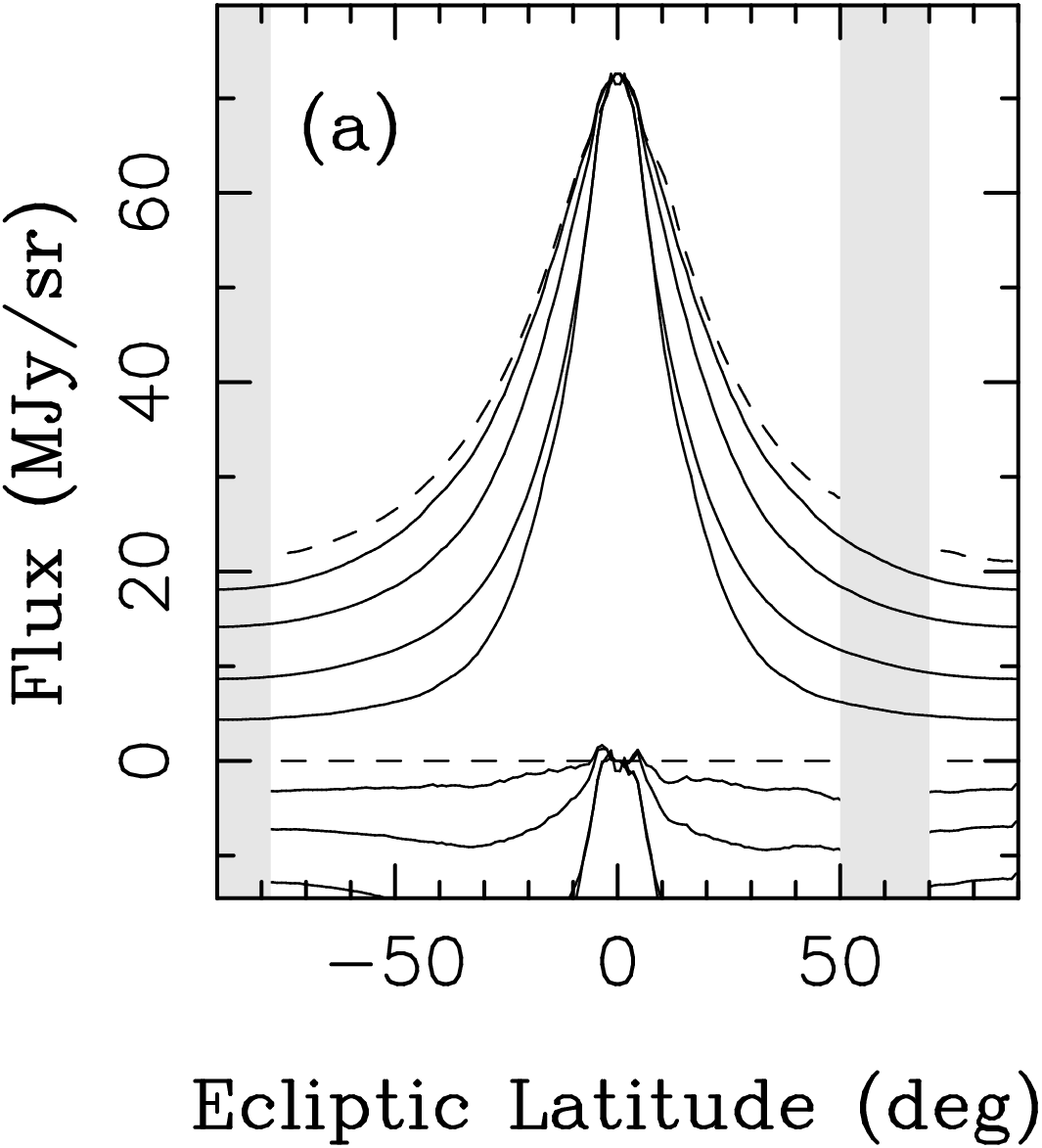}\hspace*{5.mm}
\plotone{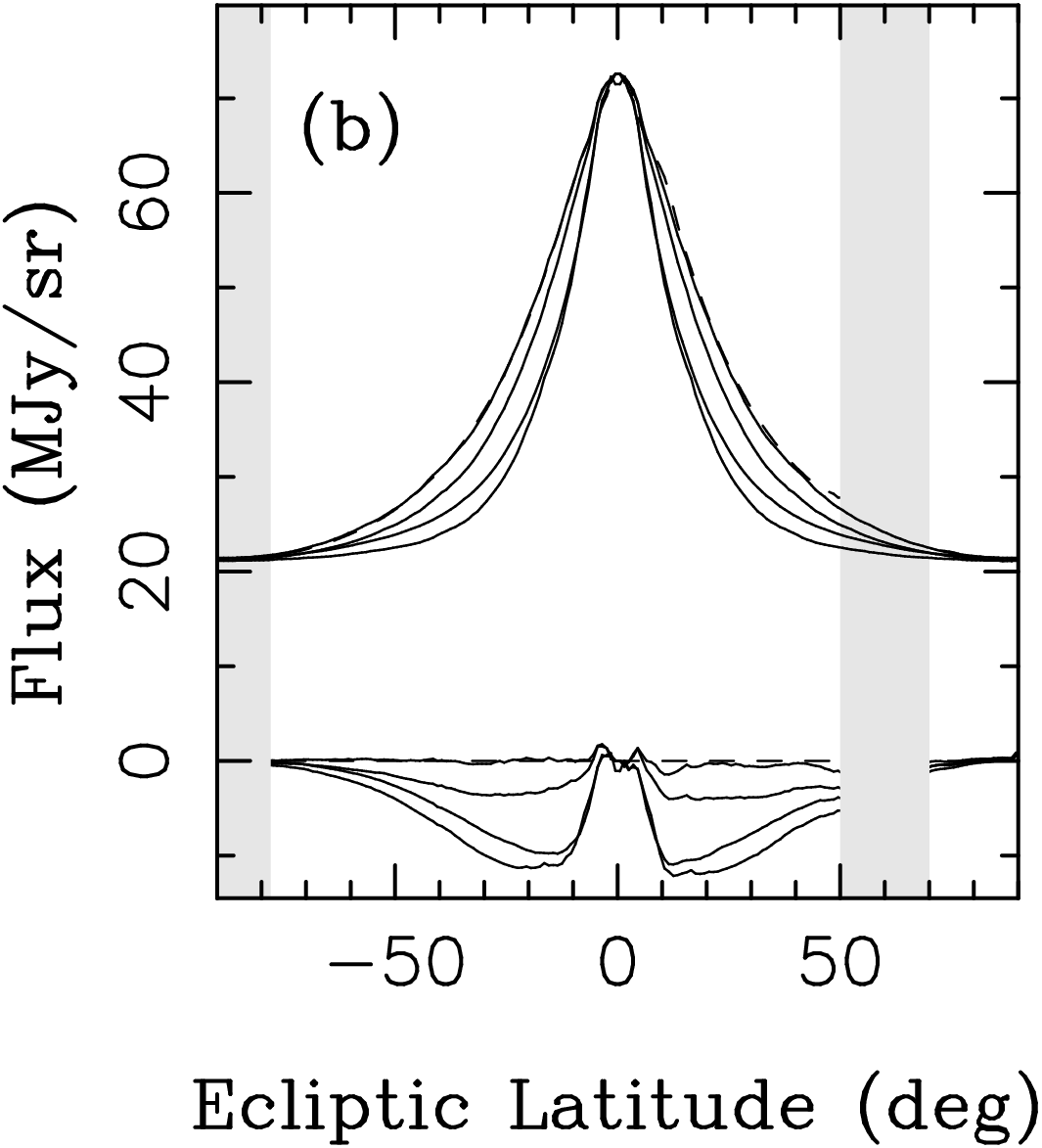}
\caption{Dependence of the shape of 25-$\mu$m profiles produced by $D=100$ $\mu$m JFC particles on 
$t_{\rm col}$. The dashed line in each panel shows the mean 25-$\mu$m IRAS profile for $l_\odot=90^\circ$. 
The upper solid curves show the model results for the same elongation. The bottom lines show the residual flux 
obtained by substracting model fluxes from the mean IRAS profile. Results for $t_{\rm col}=10^4$, 
$10^5$, $5\times10^5$ and $10^6$ yr are shown in each panel with broader profiles corresponding to 
the larger $t_{\rm col}$ values. In panel (a), we show results for the single-source model with JFC particles
only. The results of the two-source model with JFC and OCC particles are illustrated in (b). We 
included the OCC component in the model to try to compensate for the defficient polar fluxes from
JFC particles with short $t_{\rm col}$. Profiles with $t_{\rm col}\lesssim 5\times10^5$ yr do not 
match IRAS observations well.} 
\label{tcol}
\end{figure}

\begin{figure}
\epsscale{0.5}
\plotone{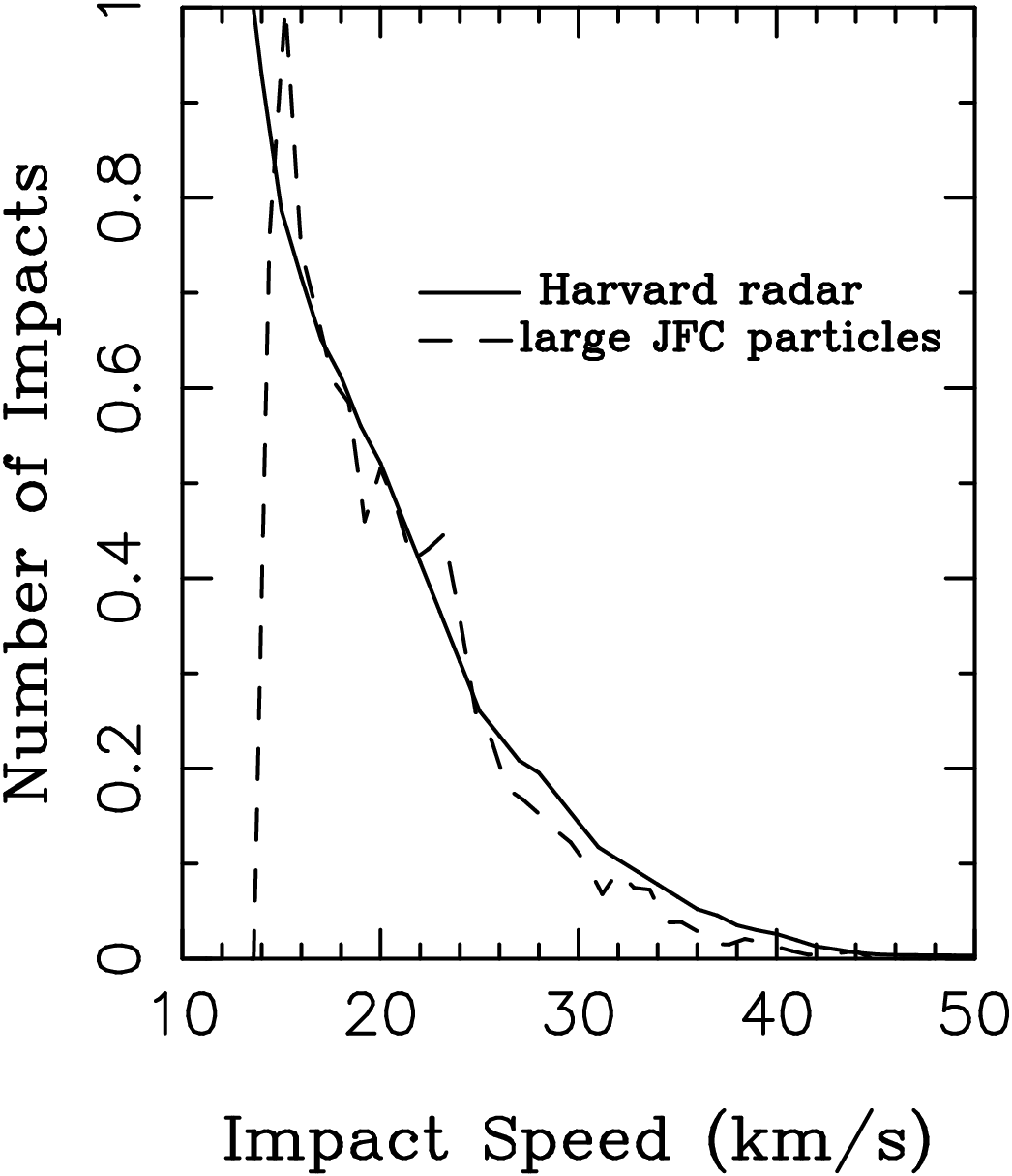}
\caption{Comparison of atmospheric entry speeds of $D=1$-mm JFC particles with $t_{\rm col}=10^{4}$ yr
with the Harvard meteor radar data (Taylor, 1995).  
There is a good agreement between the two distributions for $>$20 km s$^{-1}$. The number of impacts 
from large JFC particles drops at $<$15 km s$^{-1}$. The Harvard data is affected by strong biases 
for $<$20 km s$^{-1}$, because the detectable ionization level produced by a meteor is a strong function 
of meteor speed.} 
\label{radar}
\end{figure}

\begin{figure}
\epsscale{0.5}
\plotone{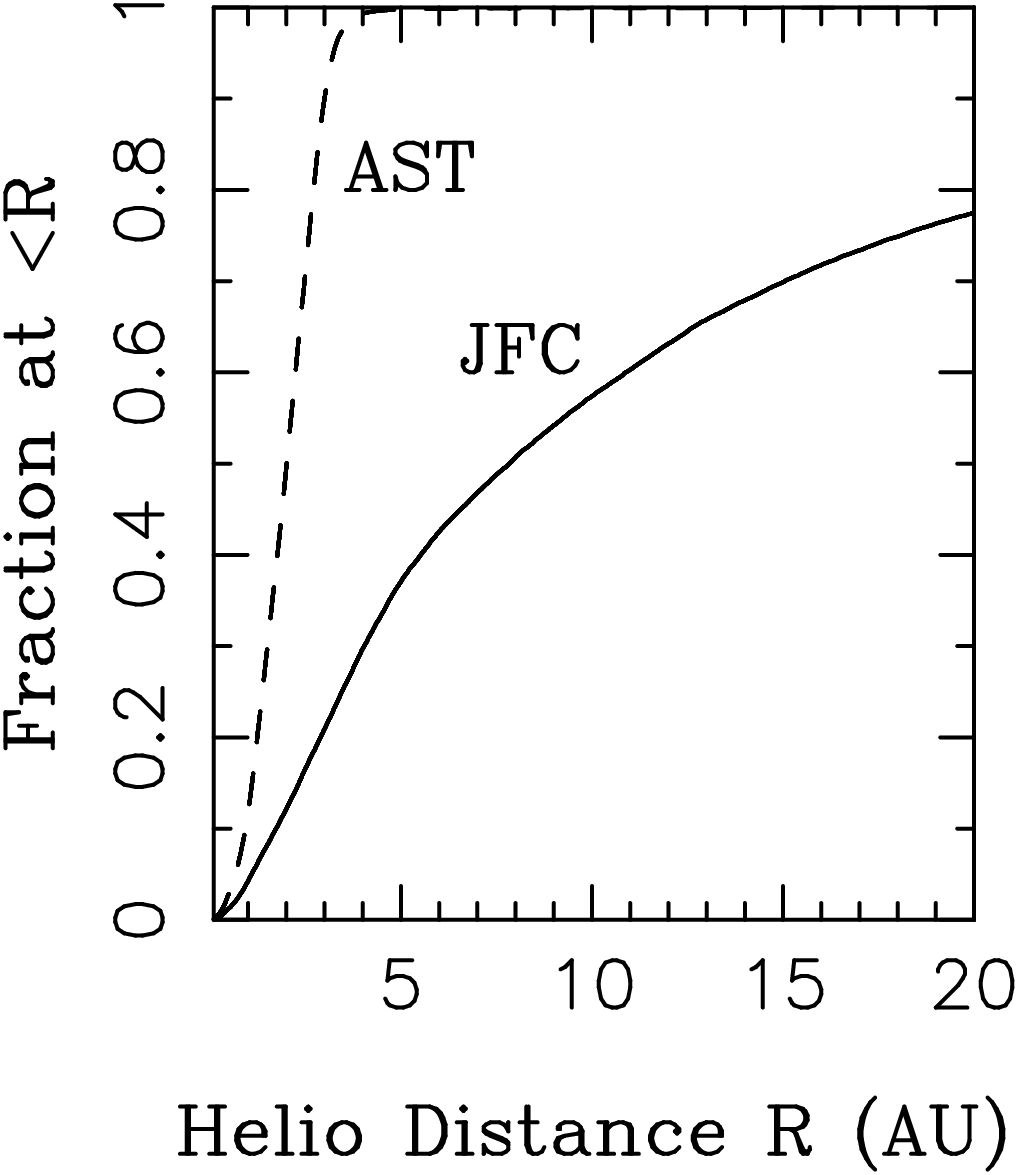}
\caption{Cumulative distribution of JFC (solid line) and asteroidal (dashed) particles as a function 
of heliocentric distance $R$. For each $R$, the value on the Y axis gives the fraction of particles
(or equivalently fraction of the total mass) contained within a sphere of radius $R$ around the Sun.
The JFC particles show a shallower slope with about 70\% having $R>4$~AU. Conversely, 99\% of 
asteroidal particles have $R<4$ AU. Note that the distributions shown here have been normalized
to 1 and do not reflect the actual relative contribution of JFC and asteroidal particles to the 
zodiacal cloud. This figure merely shows the trends in both populations with heliocentric distance.} 
\label{mass}
\end{figure}

\begin{figure}
\epsscale{0.5}
\plotone{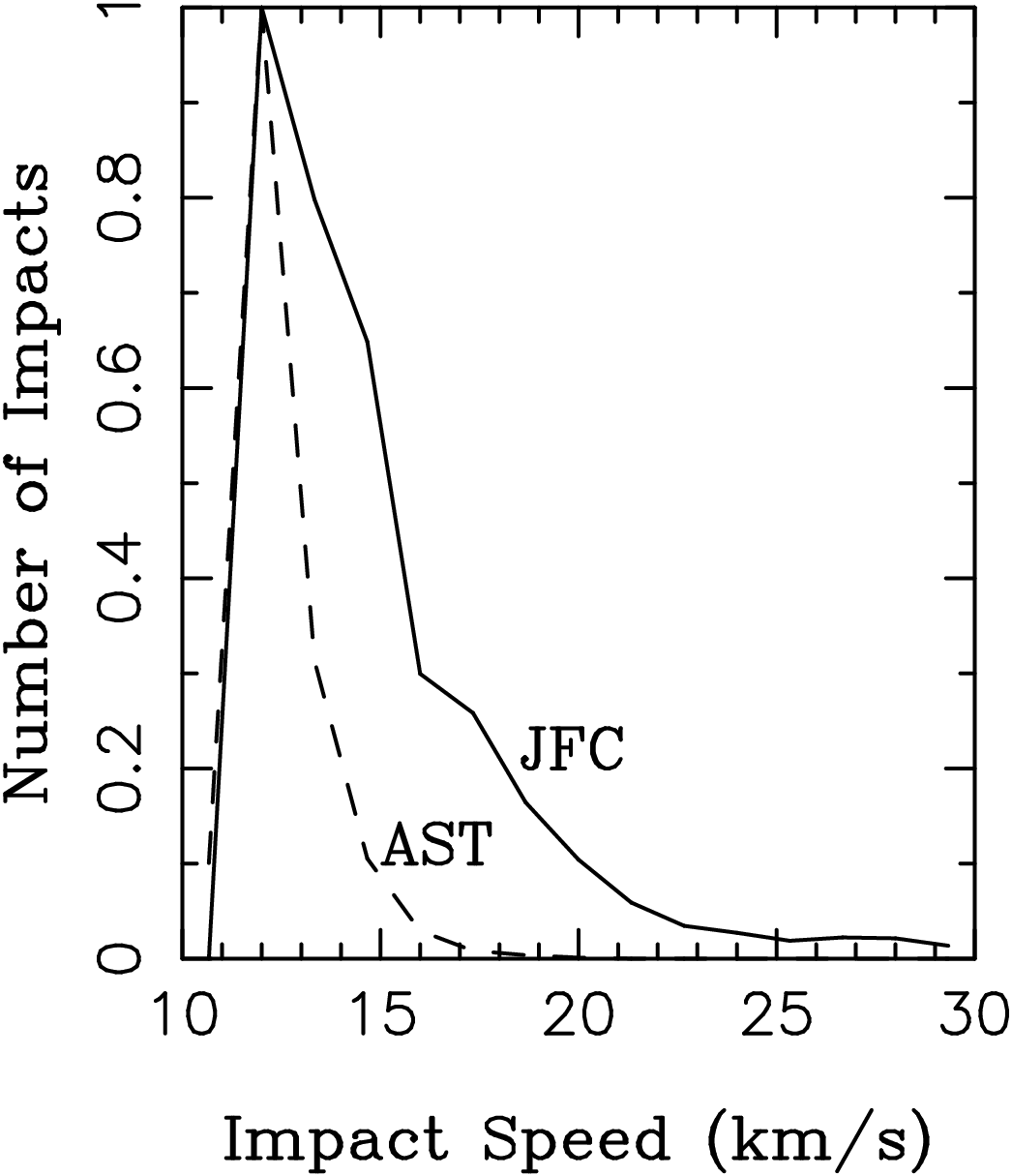}
\caption{Model distributions of Earth-impact speed of JFC (solid line) and asteroidal (dashed) particles 
with $D=200$ $\mu$m. Since the effects of the gravitational focusing have been accounted for in 
the calculation, the minimum impact speed is equal to the escape velocity from the Earth's surface, or about 
11.2 km s$^{-1}$. Majority of JFC particles have the impact speeds in the 11.2-15~km~s$^{-1}$ range. 
JFC particles with larger impact speeds have lower impact probability but are important for interpretation 
of the meteor radar data (e.g., Wiegert et al., 2009).} 
\label{disv}
\end{figure}

\begin{figure}
\epsscale{0.5}
\plotone{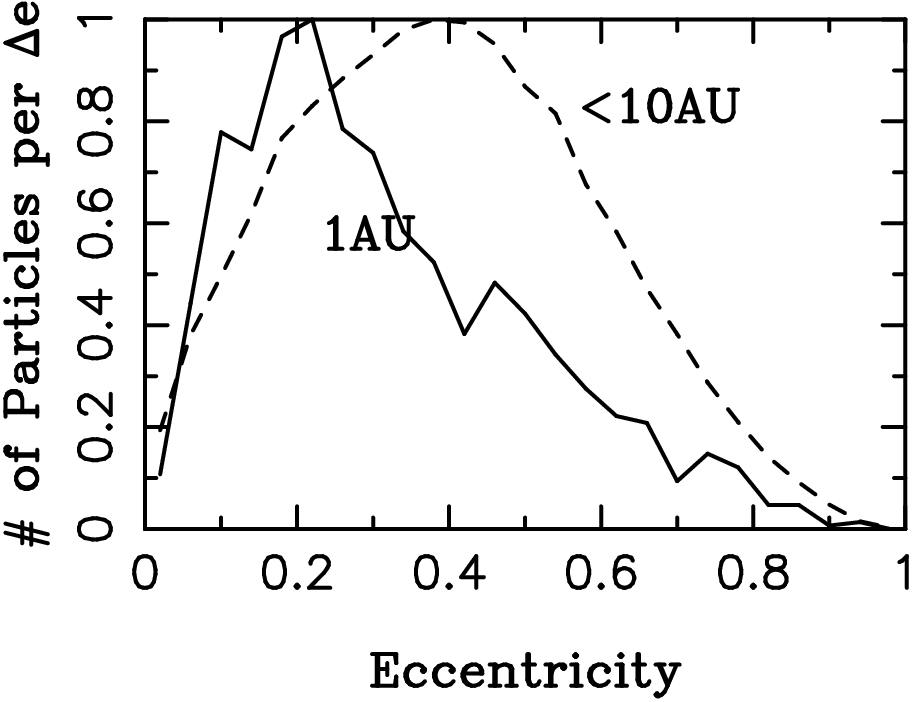}\\
\vspace*{5.mm}
\plotone{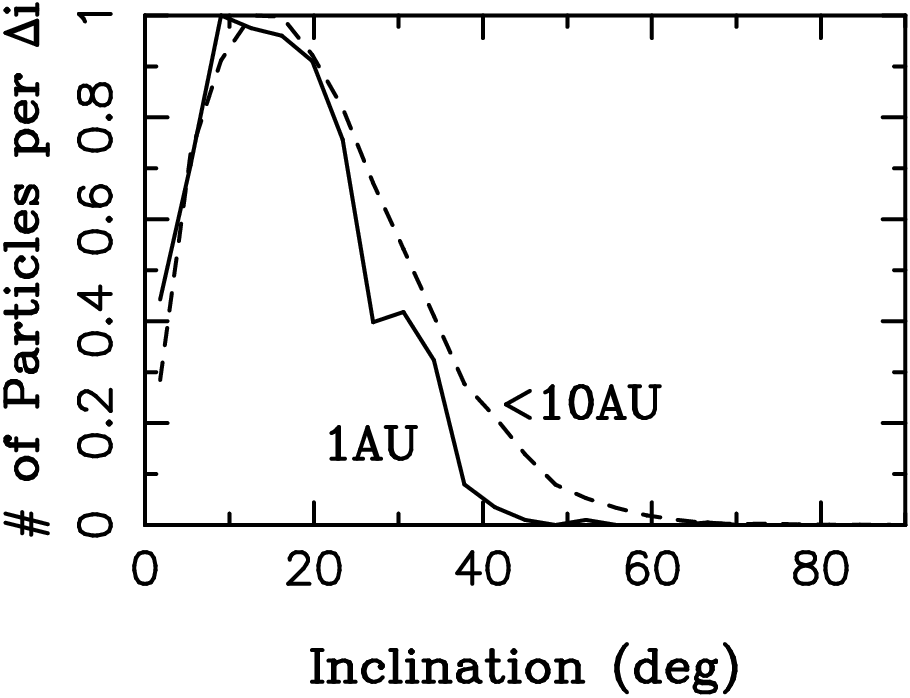}
\caption{Eccentricity (top panel) and inclination (bottom) distributions of JFC particles in our 
model. The dashed lines show the distributions for all JFC particles with $R<10$~AU. The solid
lines show the distribution for $0.9<R<1.1$ AU. The upper plot illustrates that the orbits of JFC particles 
drifting by PR drag become nearly circularized before reaching 1~AU. The inclination distribution 
does not change much during this evolution.} 
\label{orbel}
\end{figure}

\begin{figure}
\epsscale{0.7}
\plotone{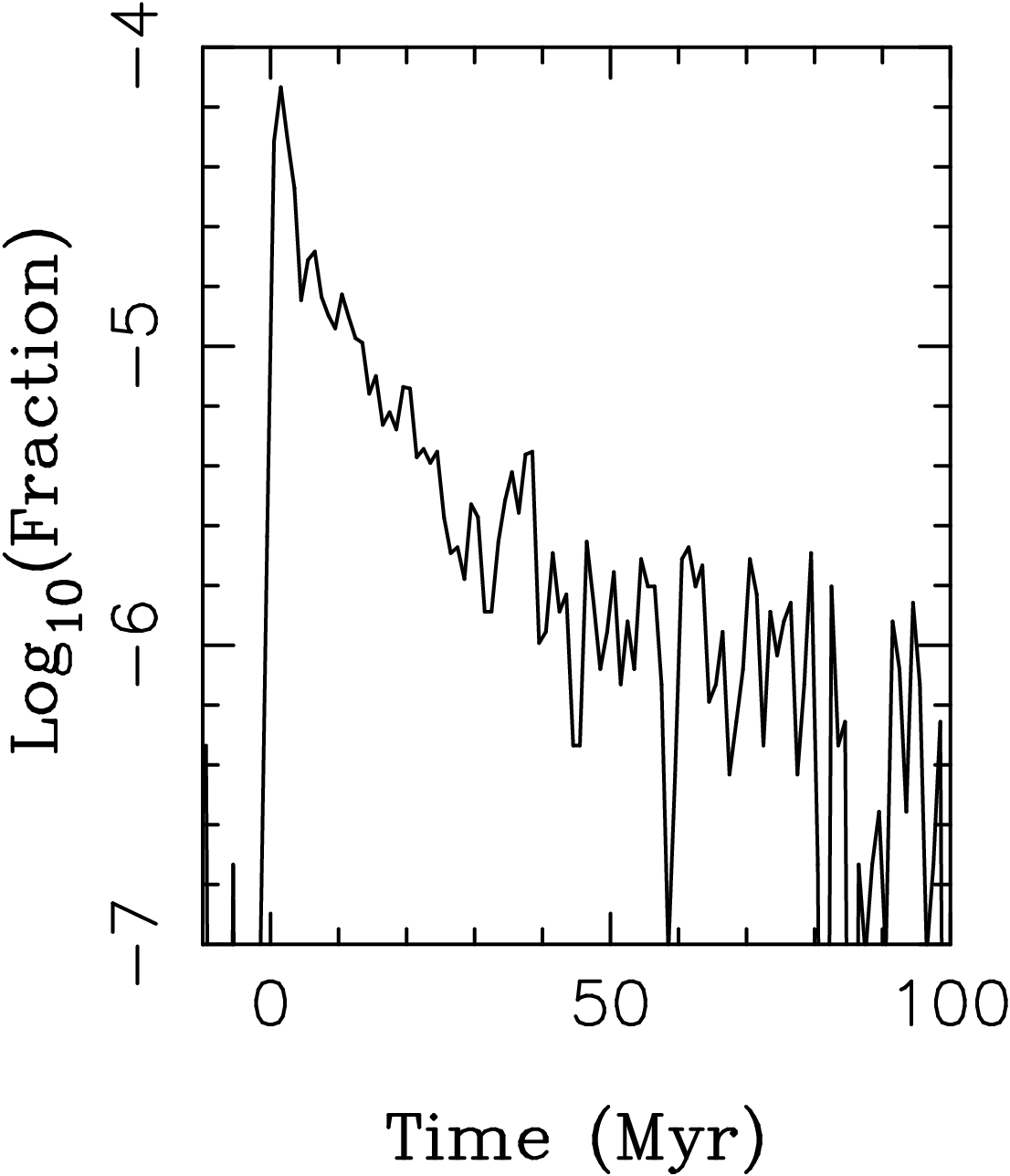}
\caption{Number of objects on JFC-like orbits during LHB as a fraction of the total number of planetesimals 
in the pre-LHB trans-planetary planetesimal disk. The fraction was determined from the n22 simulation of the 
Nice model in NV09. We extracted all orbits from that simulation with perihelion diatance $q < 2.5$ AU, 
orbital period $P<20$ yr and assumed that the physical lifetime of these objects was $10^4$ yr (LD97).
We also used an averaging window of 1 My to improve the statistics. The total mass of the JFC population can be 
estimated from this plot by multiplying the fraction shown here by the initial mass of the trans-planetary disk. 
With the 35 Earth-mass disk, the peak in the mass of the JFC population at $t \approx 0$ corresponds to 
$\sim$0.3 lunar masses.} 
\label{scat}
\end{figure}

\begin{figure}
\epsscale{0.75}
\plotone{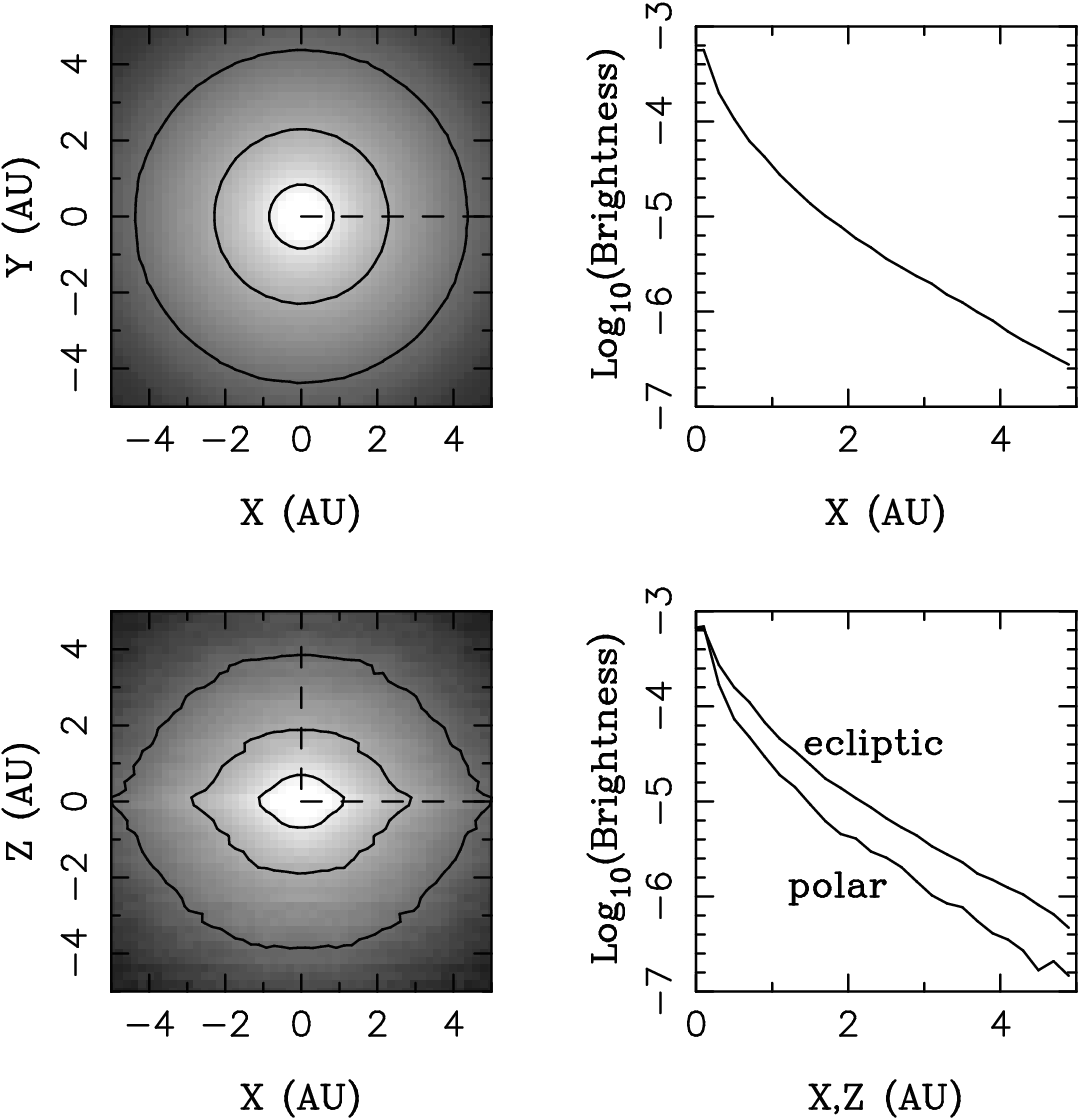}
\caption{Zodiacal cloud brightness at 24 $\mu$m as seen by an observer at 10 pc. Two projections are shown: (top) polar view 
for an observer with $Z=10$ pc; and (bottom) side view of an observer in the ecliptic plane ($Y=10$ pc). The three 
isophotes in each of the two left panels correspond to $5\times10^{-4}$, $5\times10^{-5}$ and $5\times10^{-6}$ Jy 
AU$^{-2}$ with 1 AU$^2$ at 10 pc corresponding to 0.01 arcsec$^2$. The shading scale is linear in $\log_{10}$ of brightness. 
The right panels show the brightness variation with the heliocentric distance along the cuts denoted by the dashed lines 
in the left panels. There are two lines in the bottom-right panel corresponding to the polar and ecliptic profiles.
} 
\label{zod}
\end{figure}

\begin{figure}
\epsscale{0.7}
\plotone{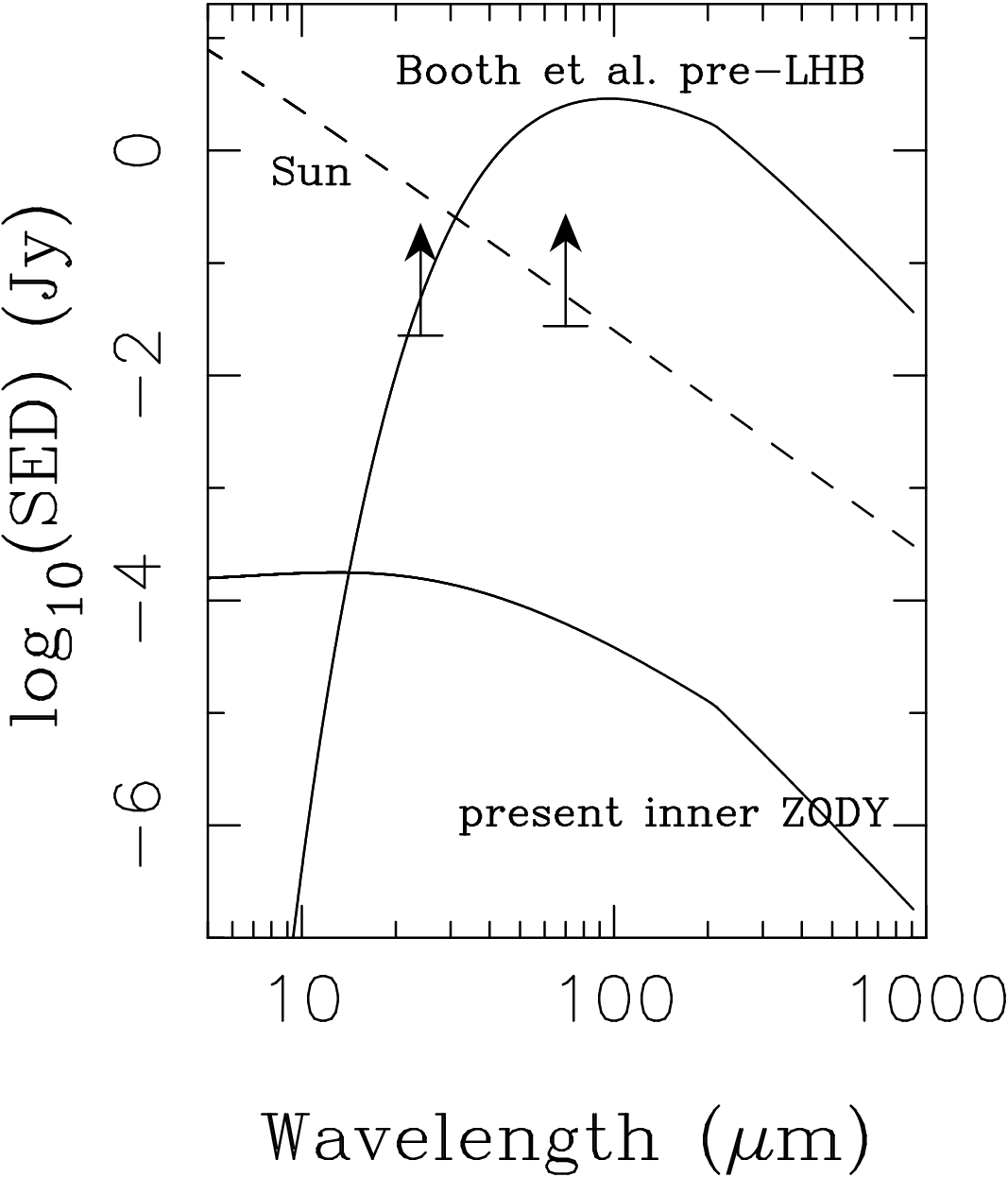}
\caption{Spectral density distribution of the present inner zodiacal cloud as seen by an observer at distance 
10 pc form the Sun. For reference, we also plot SED of the Sun and the pre-LHB trans-planetary disk as determined by 
Booth et al. (2009). The two arrows show the approximate 3$\sigma$ detection limits of the Spitzer telescope at 24 
and 70 $\mu$m (Carpenter et al., 2009; Wyatt et al., 2008).} 
\label{sed}
\end{figure}

\begin{figure}
\epsscale{0.45}
\plotone{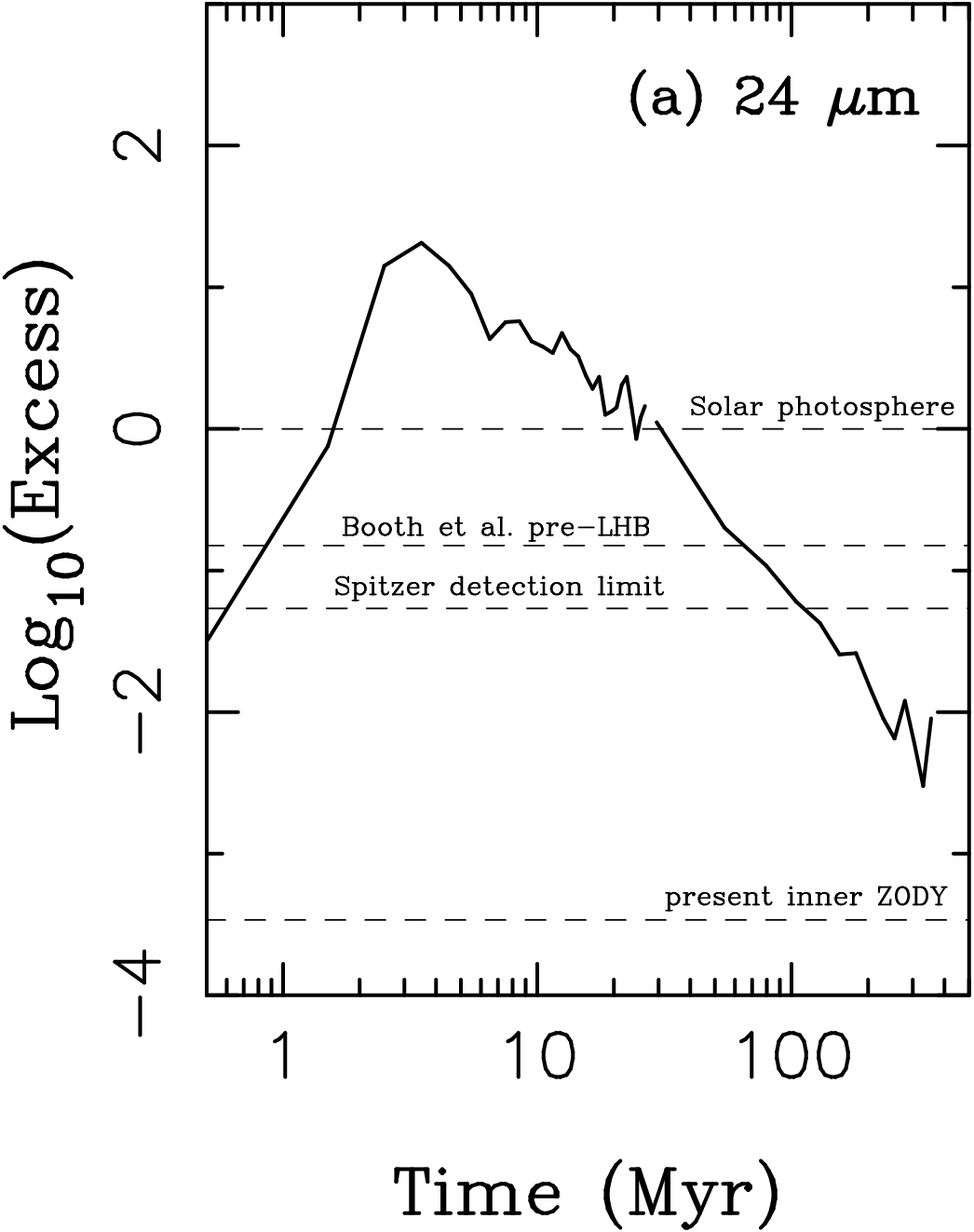}\hspace*{0.5cm}
\plotone{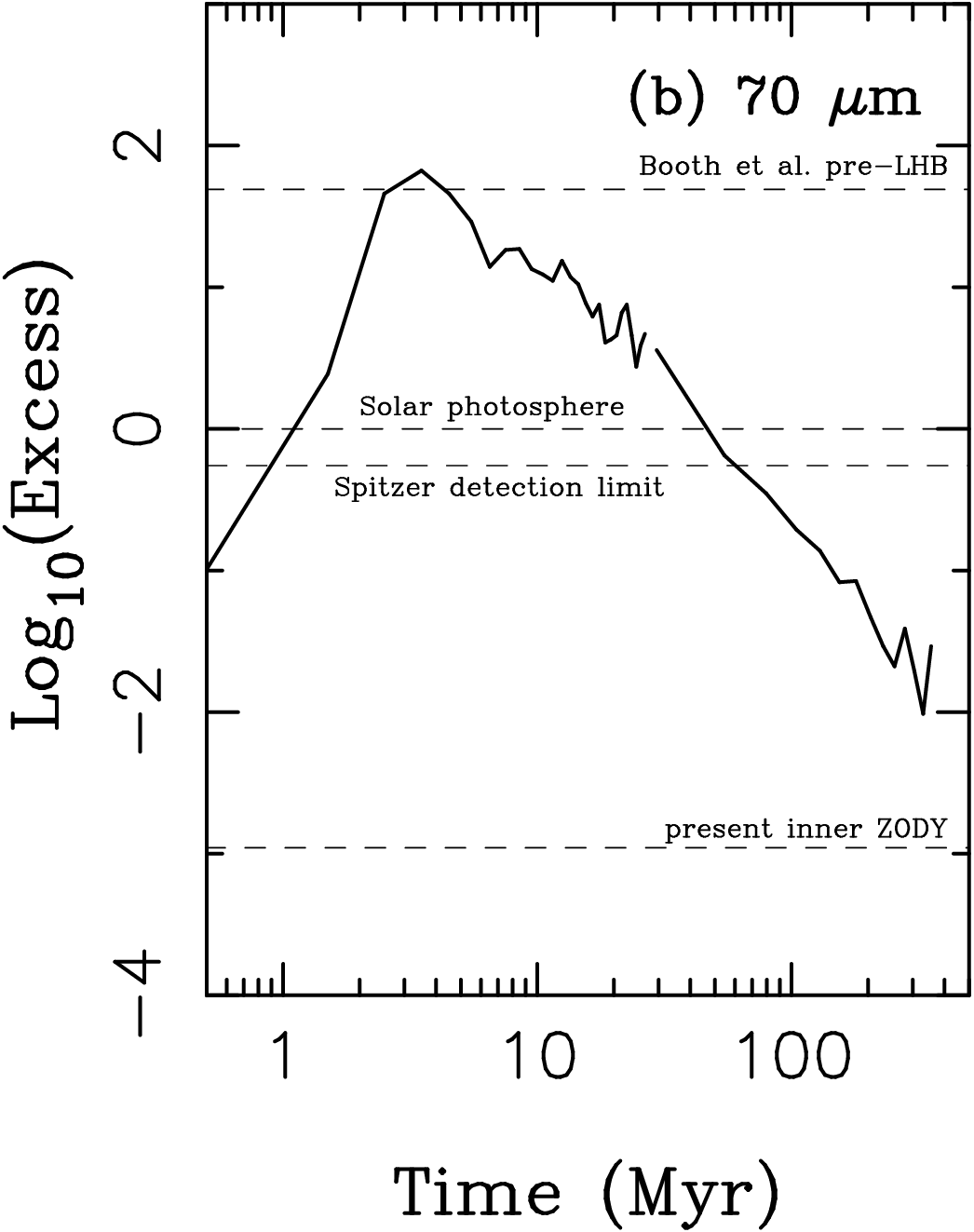}
\caption{Expected variation of excesses at 24 $\mu$m (panel a) and 70 $\mu$m (b) during LHB (solid lines). To determine 
the excess values at different times during LHB, we used Fig.~\ref{scat} to estimate the number of objects that were 
scattered from the trans-planetary disk into the JFC-like orbits. By comparing this number to the present population 
of JFCs, a scale factor has been determined to represent the brightness increase of the inner zodiacal cloud over
its current value. The discontinuity in the lines near $t=30$ Myr appears because we changed the size of the averaging 
running window, $\delta t$. For $t<30$ Myr, we used $\delta t=1$ Myr; for $t>30$ Myr, we used $\delta t=50$ Myr. The 
large $\delta t$ value is needed for $t>30$ Myr to improve the statistics. For reference, the plot also shows the values 
predicted by Booth et al. (2009) for the pre-LHB trans-planetary disk, approximate Spitzer detection limits and 
present inner zodiacal cloud (dashed lines).} 
\label{lhb}
\end{figure}

The main goal of this modeling effort was to determine the relative contribution of asteroid and cometary material 
to the zodiacal cloud. We found that asteroidal particles produced by the main belt collisions cannot produce the zodiacal cloud 
emission at large ecliptic latitudes simply because the main belt asteroids have generally small orbital inclinations, 
and because the orbital effects of planetary encounters and secular resonances at $a \lesssim 2$~AU are not powerful 
enough to spread the asteroid dust to very large orbital inclinations.
Therefore, most MIR emission from particles produced in the asteroid belt is confined to within 
$\sim$30$^\circ$ of the ecliptic (Fig. \ref{ast1}). Conversely, the zodiacal cloud has a broad latitudinal 
distribution so that strong thermal emission is observed even in the direction to the ecliptic poles (Fig. \ref{mean}). 

Based on the results discussed in \S4, we proposed that $\gtrsim$90\% of the zodiacal cloud emission at MIR wavelengths 
comes from dust grains released by Jupiter-family comets, and $\lesssim$10\% comes from the Oort cloud comets and/or 
asteroid collisions. We argued that disruptions/splitting events of JFCs are more likely to produce the bulk of observed 
dust in the inner solar system than the normal JFC activity. The relative importance of JFC and Kuiper-belt particles 
beyond Jupiter has yet to be established. 

Using our model results, we estimated the total cross-section area and mass of particles in the zodiacal cloud,  
current and historical accretion rates of dust by planets and the Moon, and discussed the implications of our work 
for studies of micrometeorites and debris disks. We found that JFC particles should dominate the terrestrial accretion 
rate of micrometeoroids. This may explain why most antarctic micrometeorites have primitive carbonaceous composition. 
If the spontaneous comet disruptions are also common in the hot exozodiacal debris disks, the collisional paradigm 
used to explain their properties may not be as universal as thought before.  

\acknowledgements
This work was supported by the NASA Planetary Geology and Geophysics and Planetary Astronomy programs. The work of DV 
was partially supported by the Czech Grant Agency (grant 205/08/0064) and the Research Program MSM0021620860 of the Czech
Ministry of Education. We thank A. Morbidelli and L. Dones for their insightful comments on the manuscript.

\clearpage

\clearpage

\clearpage

\clearpage

\clearpage

\clearpage

\clearpage

\clearpage

\clearpage

\clearpage

\clearpage

\clearpage

\clearpage

\clearpage

\clearpage

\clearpage

\clearpage

\clearpage

\clearpage

\end{document}